\documentclass[aip,amsmath,amssymb,reprint]{revtex4-1}

\usepackage{graphicx}
\usepackage{dcolumn}
\usepackage{bm}
\usepackage{svg}
\usepackage{mathtools}
\usepackage[utf8]{inputenc}
\usepackage[T1]{fontenc}
\usepackage{mathptmx}
\usepackage{etoolbox}
\usepackage{hyperref}
\usepackage{caption}
\usepackage{subcaption}

\makeatletter
\def\@email#1#2{
 \endgroup
 \patchcmd{\titleblock@produce}
  {\frontmatter@RRAPformat}
  {\frontmatter@RRAPformat{\produce@RRAP{*#1\href{mailto:#2}{#2}}}\frontmatter@RRAPformat}
  {}{}
}
\makeatother
\begin{document}

\title[]{Reaction-limited evaporation for the color-gradient lattice Boltzmann model}

\author{Gaurav Nath}
\altaffiliation{Email: n.gaurav@fz-juelich.de}
\author{Othmane Aouane}
\altaffiliation{Email: o.aouane@fz-juelich.de}
\affiliation{ 
Helmholtz Institute Erlangen-N{\"u}rnberg for Renewable Energy,
Forschungzentrum J{\"u}lich, Cauerstr. 1, 91058 
Erlangen, Germany
}
\author{Jens Harting$^{\text{1,}}$ }
\altaffiliation{Email: j.harting@fz-juelich.de}

\affiliation{ 
Department of Chemical and Biological Engineering and Department of Physics,
Friedrich-Alexander-Universit{\"a}t Erlangen-N{\"u}rnberg, Cauerstr. 1, 91058 Erlangen, Germany
}

\date{\today}

\begin{abstract}
We present a method to achieve reaction-limited evaporation for the color-gradient lattice Boltzmann multicomponent model. Our approach involves a systematic way to remove fluid mass from the interface region in order to achieve evaporation rates similar to those in a reaction-limited regime. Through various tests, our method demonstrates accurate and consistent results for different interface shapes across a wide range of evaporation flux magnitudes. A single free parameter is required to choose the evaporation sites where fluid mass is exchanged between the components. We find that at unit density ratio, this single parameter allows for the correct description of an arbitrarily shaped interface with an error of less than 5\%. For density contrasts, accurate results are observed for lower evaporation flux magnitudes and density ratios. Our proposed method can be applied to isothermal reaction-limited scenarios, such as evaporation in pure vapor or under a gas draft. It can also handle weakly space-time-dependent fluxes, making it suitable for specific non-isothermal applications such as drop evaporation from heated substrates.
\end{abstract}

\maketitle

\section{\label{sec:intro}INTRODUCTION}
Evaporation can occur in different regimes, specifically diffusion-limited (DL) and reaction-limited (RL), depending on the governing mechanism. In the diffusion-limited regime, vapor molecules move away from the liquid-vapor interface due to diffusion. This regime is commonly observed when evaporation occurs in a still mixture of gases and is often chosen for studying evaporation in droplets \cite{hu2005analysis,dunn2008mathematical, MURISIC_KONDIC_2011,stauber2014lifetimes,saenz2017dynamics,nguyen2018analytical,masoud2021evaporation,larsson2023comparison}. However, in scenarios such as evaporation in a vacuum, a pure vapor phase, or in the presence of a gas draft, the DL regime may not be applicable \cite{cazabat2010evaporation}. In such cases, the RL regime, where the transfer rate of molecules across the liquid-vapor interface is the limiting mechanism, should be considered.

RL evaporation plays a crucial role in various applications. It is considered in continuum models for drying in porous media \cite{whitaker1977simultaneous,ahmad2020non}, with applications in soil water evaporation \cite{li2019evaluation}  and water evaporation in polymer electrolyte fuel cells \cite{safi2017experimental}. RL evaporation is also utilized in studying the evaporation of molten metals, such as in the vacuum refining of Manganese (Mn) steel melts \cite{chu2021kinetic} and the vacuum evaporation of pure metals like Titanium and Zirconium at high temperatures in rarified environments \cite{poullain2022vacuum}. Additionally, in the context of drop evaporation, the RL regime is significant for the evaporation of micro-drops of pure liquids \cite{semenov2012computer} and for drop evaporation from heated substrates \cite{MURISIC_KONDIC_2011,maki2011fast,pham2017drying}.

The Hertz-Knudsen (HK) relation\cite{hertz,knudsen}, based on the kinetic theory of gases, is often used to model the local mass flux $\phi$ leaving the interface in RL evaporation. Using the Clausius-Clapeyron law, the HK relation can be expressed as
\begin{equation}
    \phi = \frac{\Lambda \, \rho_{v} \, \mathcal{L}}{T_{sat}^{3/2}} \sqrt{\frac{M_{v}}{2 \pi R_{g}}} \; \left[T_{i} - T_{sat} \right].
\end{equation}

Here, $0 \le \Lambda \le 1$ represents the accommodation coefficient (a measure of liquid volatility), $M_{v}$ is the molecular mass of the vapor, $\rho_{v}$ is the density of the vapor, $R_{g}$ is the universal gas constant, $\mathcal{L}$ is the latent heat of vaporization, $T_{i}$ is the interface temperature (assumed to be continuous across the interface), and $T_{sat}$ is the saturation temperature. The difference between $T_{sat}$ and $T_{i}$ drives the flux. 
Despite known limitations \cite {persad2016expressions}, the HK relation's simplicity has led to extensive use in theoretical works on droplet evaporation \cite{ajaev2005spreading,sodtke2008dynamics, MURISIC_KONDIC_2011,maki2011fast,pham2017drying,larsson2023comparison}. It has been adopted for the non-equilibrium one-sided (NEOS) model for evaporation due to its capability of decoupling the drop dynamics from the vapor side \cite{murisic2008modeling}. In this work, we adopt a similar approach for Lattice Boltzmann (LB) simulations.

Among the different classes of LB multicomponent models\cite{liu2016multiphase}, the color-gradient (CG) approach\cite{gunstensen1991lattice,gunstensen1992microscopic} has found widespread application in studying two-phase flows in porous media\cite{zhao2020simulation,diao2021numerical,lin2021spontaneous} and thermocapillary flows\cite{fu2021numerical,fu2022numerical,fu2023three}. Its strength lies in achieving large kinematic viscosity ratios (up to 1000)\cite{liu2016multiphase} and independent tuning of crucial parameters like surface tension and interface thickness \cite{leclaire2017generalized}. In recent years, several works\cite{leclaire2017generalized,wen2019improved,lafarge2021improved,subhedar2022color,mora2021optimal} have improved the CG model, making it viable for general three-dimensional multiphase/multicomponent flows and overcoming the limitations of the previous implementations. However, unlike the pseudo-potential\cite{shan1993lattice} (PP) and free-energy\cite{swift1996lattice} (FE) based multicomponent models, the CG model does not have an inherent method of achieving evaporation. The multicomponent PP and FE methods are well-known for exhibiting diffusion-limited evaporation\cite{hessling2017diffusion,ledesma2014lattice} and subsequently used for drying applications\cite{XH18a,XH19,fei2022pore,fei2024pore,zhao2017modeling,zhao2019harnessing}. The CG model does not exhibit similar diffusion behavior as it is intended for immiscible fluids.

Additionally, the CG model uses an isothermal equation of state that does not account for phase change. Aursj\o $\,$ and Pride\cite{aursjo2015lattice} proposed an LB model for two partially miscible fluids where the fluid has two distinct regions - an interfacial region where color separation and surface tension are enforced and a miscible region where an advection-diffusion equation governs concentrations. Subsequently, they demonstrated the dissolution of one fluid component into another while maintaining a freely moving interface. The model shares CG features and could be adopted for diffusion-limited evaporation but introduces the complication of solving two different algorithms in complementary fluid regions (interface and bulk). Another approach used in the CG literature\cite{scheel2024enhancement,liu2021pore} is to convert one fluid component into the other (red to blue) at the domain's boundaries. However, the said conversion cannot be considered evaporation as it does not obey any evaporation regimes. Considering the above literature, there is a need for a straightforward evaporation method in the CG model. Considering the limitations of the CG model with diffusion-limited evaporation, we propose a way to achieve reaction-limited evaporation at minimum computational overhead with no change to the core algorithm.

In RL evaporation, the rate of mass loss is directly related to the liquid-vapor surface area. The proposed method involves removing fluid mass from the lattice sites in the interface region formed between the fluid components. We have developed a consistent way of selecting these evaporation sites based on the calculated color-gradient magnitude and a threshold value (free parameter). The threshold, which depends on a given set of CG model parameters, is chosen through a series of benchmarks. As a result, the current method can yield results consistent with RL evaporation. 

The remaining sections of the paper are organized as follows: Section \ref{sec:method} explains the CG model and the RL evaporation algorithm, while Section \ref{sec:results} presents benchmarks and discusses the strategy for choosing the free parameter. Finally, Section \ref{sec:end} summarizes the findings and explores potential use cases of the current method.

\section{SIMULATION METHOD} \label{sec:method}

\subsection{Color-gradient lattice Boltzmann multicomponent model}
In this work, we adopt the generalized three-dimensional color-gradient (CG) model proposed by Leclaire et al. \cite{leclaire2017generalized}. For a given fluid component $k$, $f_{i}^{k}(\bm{x},t)$ represents the density of particles at site $\bm{x}$ at time $t$ moving along the link direction $i$ with velocity $\bm{c}_{i}$. A \textit{color-blind distribution function} is then defined as $ f_{i} = \sum_{k} f_{i}^{k} $ with the component density, total fluid density and color-blind population velocity given by $\rho_{k} = \sum_{i} f_{i}^{k}$,  $\rho = \sum_{k} \rho_{k}$ and $\bm{u} = \frac{1}{\rho} \sum_{i} f_{i}\bm{c}_{i}$, respectively. The collision step is performed on the color-blind population via the BGK \cite{bhatnagar1954model} (Bhatnagar-Gross-Krook) single relaxation time collision operator ($\Omega_{BGK}$), given by
\begin{equation}
    \Omega_{BGK} (f_{i}) = f_{i} -\frac{\Delta t}{\tau}[f_{i} - f_{i}^{eq}]. \label{eq:collision}
\end{equation}
Here, $\tau$ is the effective relaxation time determined from the harmonic density weighted average of the component relaxation times ($\tau_{k}$) as $\frac{1}{\tau} = \sum_{k} \frac{\rho_{k}}{\rho} \frac{1}{\tau_{k}}$. The component viscosity $\nu_{k}$ is defined as $\nu_{k} =  c_{s}^{2} (\tau_{k} - \frac{\Delta t}{2})$, where $c_{s}$ is the lattice speed of sound and the equilibrium distribution function $f_{i}^{eq}$, based on a Maxwellian distribution, is given by
\begin{eqnarray}
    && f_{i}^{eq} (\rho,\bm{u}) = \nonumber \\ && \rho \left[A_{i} + B_{i} \bar{\alpha} + W_{i} \left(3 (\bm{c_{i}}\cdot\bm{u}) + \frac{9}{2} (\bm{c_{i}}\cdot\bm{u})^{2} - \frac{3}{2} (\bm{u}\cdot\bm{u}) \right) \right].  \label{eq:feq}
\end{eqnarray}
$A_{i}$, $B_{i}$, and  $W_{i}$ are lattice-specific weights and can be found in work \cite{leclaire2017generalized}. The parameter $\bar{\alpha}$ is introduced to account for the density ratio ($\gamma$) between fluid components and is determined using the arithmetic density weighted average of the component free parameters ($\alpha_{k}$) as $\bar{\alpha} = \sum_{k} \frac{\rho_{k}}{\rho} \alpha_{k}$. The least dense fluid component is assigned $\alpha_{k} = W_{0}$ and $\alpha_{k} = 1 - \frac{[1 - W_{0}]}{\gamma_{k}}$ is used for the rest, where $\gamma_{k} = \rho_{k}^{in} / \rho_{min}^{in}$ is the ratio of the initial densities of a given component $k$ to the least dense component. The pressure in each homogeneous region is then given by
 \begin{equation}
     p_{k} = \rho_{k} \zeta [1 - \alpha_{k}], \label{eq:EOS}
 \end{equation}
  where $\zeta$ is a lattice dependent weight\cite{leclaire2017generalized}, related to the isothermal speed of sound $(c_{s}^{k})^{2}$.

 A perturbation operator ($\Omega_{pert}$) \cite{reis2007lattice,gunstensen1991lattice} is used to enforce surface tension in the CG model. It is given by
 \begin{equation}
     \Omega_{pert} (f_{i}) = f_{i} + \sum_{r}  \sum_{b = r + 1} \frac{9}{4} \frac{\sigma_{rb}}{\tau} \, | \bm{F}_{rb} | \, \left[W_{i} \frac{(\bm{F}_{rb} . \bm{c}_{i})^{2}}{|\bm{F}_{rb}|^{2}} - C_{i}\right], \label{eq:perturb}
 \end{equation}
where $C_{i}$ is a lattice dependent weight \cite{leclaire2017generalized}. For a given pair of fluid components $(\rho_r,\rho_b)$ having surface tension $\sigma_{rb}$, $\bm{F}_{rb}$ is the color-gradient defined as
\begin{equation}
    \bm{F}_{rb} = \nabla \left(\frac{\rho_{r} - \rho_{b}}{\rho_{r} + \rho_{b}} \right). \label{eq:cg}
\end{equation}
A recoloring operator $(\Omega_{recol})$ is then used to recover the component populations from the color-blind population 
\begin{equation}
    f_{i}^{r*} \equiv \Omega_{recol}^{r}(f_{i})  = \frac{\rho_{r}}{\rho}  f_{i} +   \sum_{\substack{b \\ b \neq r}} \beta \frac{\rho_{r}\rho_{b}}{\rho^{2}} \,cos(\vartheta_{rb}) \, f_{i}^{eq}(\rho,0), \label{eq:recolor}
\end{equation}
where, $0 \le \beta \le 1$ controls the interface thickness and $\vartheta_{rb}$ is the angle between the color-gradient vector $\bm{F}_{rb}$ and lattice velocity vector $\bm{c}_{i}$. 

After the above operations (collision $\rightarrow$ perturbation $\rightarrow$ recoloring), the resulting distribution functions $f_{i}^{k*}$ are then streamed to the neighboring lattice nodes, completing one timestep
\begin{equation}
        f_{i}^{k}(\bm{x}+\bm{c}_{i}\Delta t, t + \Delta t) = f_{i}^{k*} (\bm{x},t). \label{eq:Stream}
\end{equation}

We use a D3Q19 lattice in the current work ($i= 0 - 18$) and the lattice spacing $\Delta x$ and timestep $\Delta t$ are taken to be unity in respective lattice units.

The gradient of the color function $\rho^{N} = \left(\frac{\rho_{r}-\rho_{b}}{\rho_{r}+\rho_{b}}\right)$ is calculated via a compact finite difference scheme, given as
\begin{equation}
    \frac{\partial \rho^{N}(\bm{x})}{\partial x_{\alpha}} = \frac{1}{c_{s}^{2}} \sum_{i} W_{i} \; \rho^{N}(\bm{x}+\bm{c}_{i}) \; \bm{c}_{i\alpha}.
\end{equation}

\subsection{Reaction-limited evaporation} \label{sec:rl_model}
In reaction-limited evaporation, the total mass loss rate ($dM/dt$) is directly proportional to the surface area of the evaporating liquid, denoted as $A_{S}$, and can be expressed as
\begin{equation}
    \frac{dM}{dt} = -\iint_{A_{S}} \phi \, dA_{S}, \label{eq:surface_flux}
\end{equation}
where $\phi$ represents the scalar mass flux normal to the elemental area $dA_{S}$, and is considered a constant in this context. 
 
In lattice Boltzmann (LB) simulations, mass removal must be accounted for at the interface region between fluid components, referred to as evaporation sites ($\bm{x} \in \bm{x}_{I} $). At each discrete evaporation site, an elementary volume of $(\Delta x)^3$, where $\Delta x$ is the lattice spacing, can be assumed. The combined mass flux leaving through its faces can be set as $\phi$. This allows for the re-expression of Eq. \ref{eq:surface_flux} for discrete evaporation sites in LB as
\begin{equation}
    \frac{dM}{dt} = -\iint_{A_{S}} \phi \, dA_{S} = - S \sum_{\bm{x} \in \bm{x}_{I}} \phi (\Delta x)^{2}.  \label{eq:satisfy}
\end{equation}
Here, $S$ represents a correction factor (a real-valued constant) introduced to account for errors resulting from the discretization of the fluid interface on a lattice, which will be further discussed later. The total mass change can also be attributed to a mass sink with rate $\varphi = d\rho / dt$ (a constant) in each elementary volume, leading to the following relations:
\begin{equation}
    \frac{dM}{dt} = - S \sum_{\bm{x} \in \bm{x}_{I}} \phi (\Delta x)^{2} = -\sum_{\bm{x} \in \bm{x}_{I}} \varphi (\Delta x)^{3}, 
\end{equation}
\begin{eqnarray}
    \phi &=& \bar{S} \, \varphi \, \Delta x  \mspace{20mu} \left[ \bar{S} = \frac{1}{S} \right].  \label{eq:relation}
\end{eqnarray}
Consequently, achieving an evaporation rate similar to reaction-limited evaporation with mass flux $\phi$ can be realized by removing fluid mass from evaporation sites at a rate of $\varphi$, following the relation in Eq. \ref{eq:relation}. In LB, $\varphi \Delta t$ represents the fluid mass to be removed per unit volume of an evaporation site in one timestep $\Delta t$. The mass exchange between the resting populations of the evaporating and surrounding fluid components is described as follows:
\begin{equation}
\label{eq:exchange}
\begin{aligned}[c]
 f_{0}^{r} (\bm{x},t)^{\text{new}} &= f_{0}^{r} (\bm{x},t) - \; \varphi \Delta t \\
 f_{0}^{b} (\bm{x},t)^{\text{new}} &= f_{0}^{b} (\bm{x},t) + \; \varphi \Delta t  
\end{aligned}
\qquad
\begin{aligned}[c]
\forall \; \bm{x} \in \bm{x}_{I}
\end{aligned}
\end{equation}
The above equation adjusts the individual component densities ($\rho_{r},\rho_{b}$) while the total density ($\rho$) at a site remains unaffected. Consequently, the method is globally mass-conserving and does not affect momentum.

\begin{figure*}
\centering
     \begin{subfigure}[b]{0.45\textwidth}
         \centering
         \includegraphics[width=\textwidth]{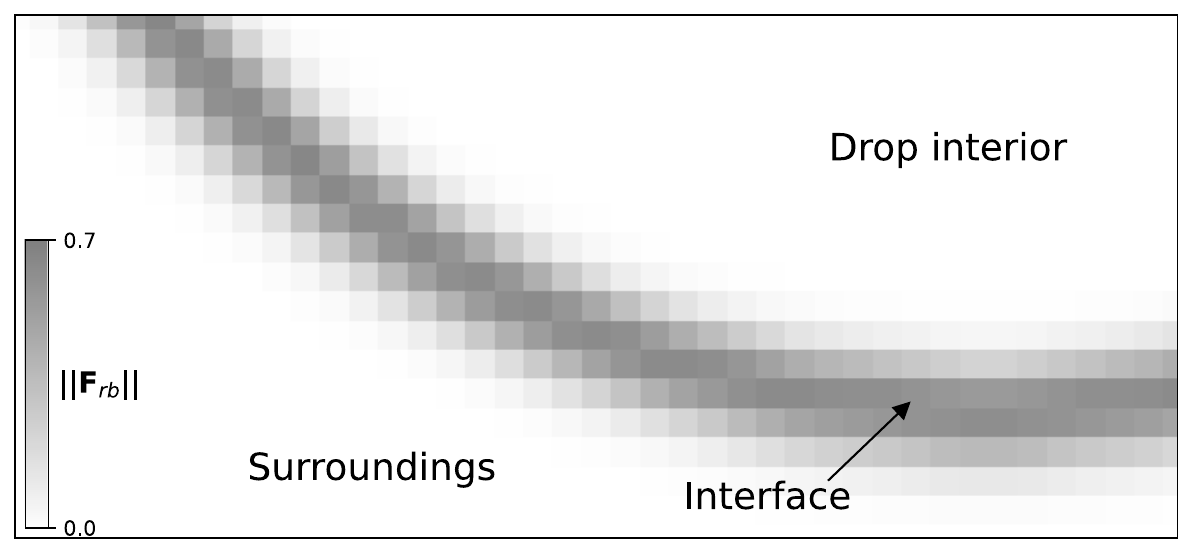}
         \caption{}
         \label{fig:S_schematic}
     \end{subfigure}
     \begin{subfigure}[b]{0.45\textwidth}
         \centering
         \includegraphics[width=\textwidth]{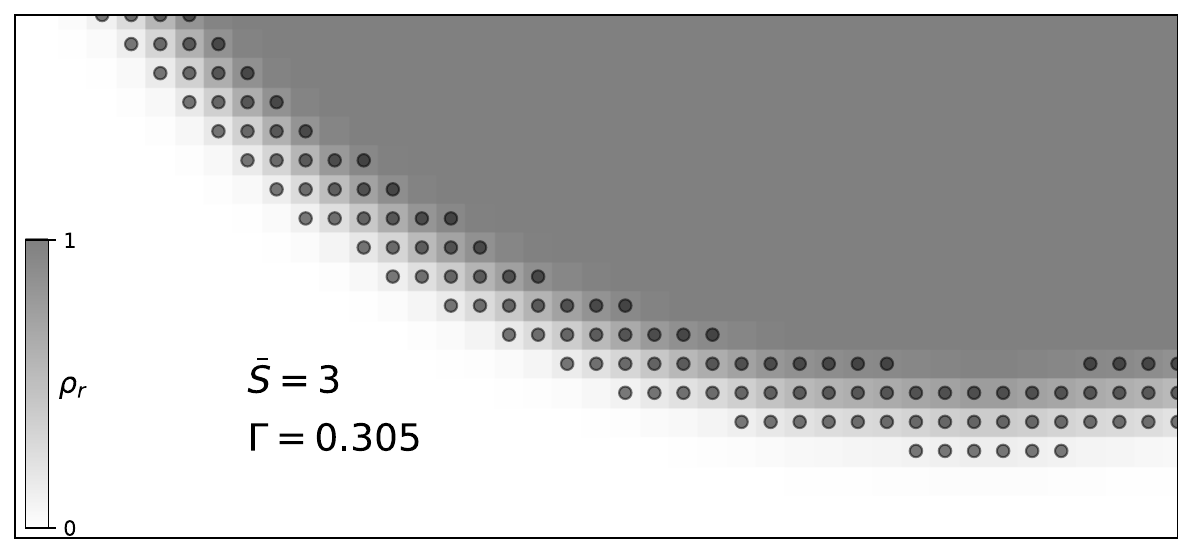}
         \caption{}
         \label{fig:S_3}
     \end{subfigure}
     \begin{subfigure}[b]{0.45\textwidth}
         \centering
         \includegraphics[width=\textwidth]{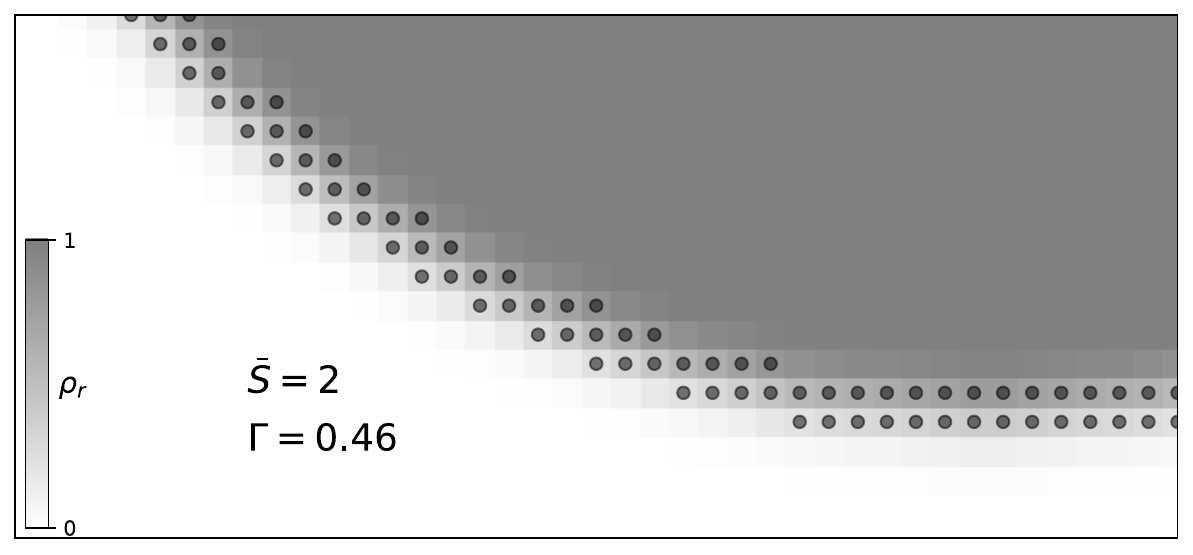}
         \caption{}
         \label{fig:S_2}
     \end{subfigure}
     \begin{subfigure}[b]{0.45\textwidth}
         \centering
         \includegraphics[width=\textwidth]{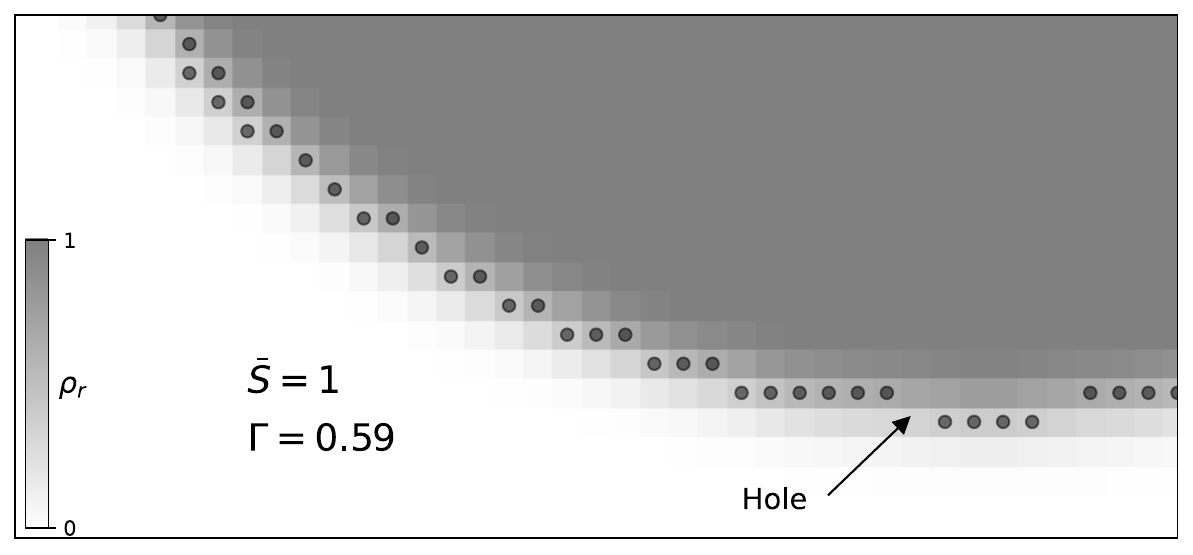}
         \caption{}
         \label{fig:S_fail}
     \end{subfigure}
\caption{ (a) Color-gradient magnitude $||\bm{F}_{rb}||$ at different lattice sites for a spherical drop. (b), (c) and (d) Evaporation sites (marked by $\bullet$) selected at the spherical interface for different $\bar{S}$ and corresponding $\Gamma$ (in lattice units). The colors show the density of the evaporating fluid $\rho_{r}$. The results shown are for unit density ratio and $\beta=0.99$.}
\end{figure*}

The sites where evaporation occurs ($\bm{x} \in \bm{x}_{I} $) can be determined using the color-gradient magnitude $||\bm{F}_{rb}||$ between two fluid components. This magnitude varies from zero in bulk to a maximum value near the interface region, as illustrated in Fig. \ref{fig:S_schematic}. A threshold ($\Gamma$) can then be applied to select evaporation sites at the interface based on the criterion 
\begin{equation}
    ||\bm{F}_{rb}|| > \Gamma.
\end{equation}
Here, $\Gamma$ is a specified constant set at the beginning of the simulation and remains unchanged throughout.
The selection of $\Gamma$ determines the value of the real-valued correction factor $\bar{S} = 1/S$. For simplicity, $\bar{S}$ is assumed to be a natural number constant. The optimal value of $\Gamma$ is determined by ensuring that the total mass loss rate is equal in both the LB simulation and the analytical solution, i.e., Eq. \ref{eq:satisfy} is satisfied. The position of the interface during evaporation is compared between the LB simulations and analytical solutions. Figures \ref{fig:S_3}, \ref{fig:S_2}, and \ref{fig:S_fail} display the evaporation sites selected at the fluid interface for $\bar{S} = 3$, $2$, and $1$, along with their corresponding $\Gamma$, for an evaporating spherical drop with unit density ratio and $\beta = 0.99$.

From the figures, it can be inferred that $\bar{S}$ roughly represents the number of layers of evaporation sites selected at the interface. However, if $\bar{S}$ is not carefully chosen, holes may appear in the selected layer, as shown in Fig. \ref{fig:S_fail} for $\bar{S}=1$. To avoid this issue, it is recommended to choose $\bar{S}>1$.

The overall method is summarized here:
\begin{enumerate}
    \item Given a flux $\phi$ and a chosen $\bar{S}$, the mass sink rate $\varphi$ is determined using Eq. \ref{eq:relation}. 
    \item The evaporation sites ($\bm{x} \in \bm{x}_{I} $) for a given pair of fluid components are chosen using the criterion $ ||\bm{F}_{rb}|| > \Gamma$. 
    \item Mass is exchanged between the fluid components at the evaporation sites using Eqs. \ref{eq:exchange}. 
    \item The color-blind distribution ($f_{i} = \sum_{k} f_{i}^{k}$) and macroscopic variables ($\rho_{k} = \sum_{i} f_{i}^{k}$,  $\rho = \sum_{k} \rho_{k}$ and $\bm{u} = \frac{1}{\rho} \sum_{i} f_{i}c_{i}$) are determined. 
    \item Collision, perturbation, and recoloring operations are systematically carried out as per Eqs. \ref {eq:collision}, \ref {eq:perturb} and \ref {eq:recolor}, respectively. 
    \item A streaming step is performed via Eq. \ref{eq:Stream}, completing one timestep. 
\end{enumerate}
The color-gradient magnitude $ ||\bm{F}_{rb}|| $ used in Step 2 uses the value determined from the previous timestep during the perturbation step to save computational effort. It is evident that this procedure has a small computational overhead and does not change the core algorithm of the CG-LB method.

\section{RESULTS AND DISCUSSION}\label{sec:results}
We demonstrate RL evaporation using the method introduced in the previous section. During evaporation, we compare the interface positions in LB simulations with analytical solutions whenever possible. The simulations involve initializing a pair of fluid components (red-blue) and achieving equilibrium based on a domain-wide velocity-based convergence criterion of $10^{-7}$. Subsequently, evaporation begins, with the red fluid assumed to be the evaporating component. The interface position is considered to be where the density of the evaporating fluid is half of the maximum value. No-slip boundary conditions are enforced on the walls using halfway-bounceback boundary conditions (BC) \cite{ladd1994numerical}. Additionally, the walls are neutrally wetting. The units for quantities such as length, velocity, $\rho$, $\sigma$, $\tau$ as well as $\phi$ and $\Gamma$, are specified in their respective lattice units. The simulations are conducted with no density contrast ($\rho_{r} = \rho_{b} = 1$), interface thickness parameter $\beta = 0.99$, surface tension $\sigma = 0.1$, and unit relaxation parameters ($\tau_{r} = \tau_{b} = 1$), unless stated otherwise.

In Section \ref{sec:shape}, the effect of interface shape on the choice of $\Gamma$ is shown for a given set of CG parameters and over a wide range of $\phi$. Section \ref{sec:cg_parms} investigates the effect of the CG model parameters on the choice of $\Gamma$. Section \ref{sec:var_flux} shows the accuracy of the current method in the case of space and time-varying evaporative flux. In Section \ref{sec:strategy}, we discuss the findings and comment on the overall method. All results are shown henceforth with $\bar{S}=3$.

\subsection{Effect of interface shape} \label{sec:shape}
\subsubsection{Flat interface} \label{sec:flat}
We consider a pseudo-one-dimensional domain with dimensions $[x,y,z] = 4 \times 128 \times 4$ , where there are no-slip walls at the boundaries in the $y$-direction and periodic boundary conditions (BC) at the remaining boundaries. The simulation begins with red fluid up to $y = 80$  and the remaining space filled with blue fluid in the $y$-direction. For a constant mass flux $\phi$ leaving through surface area $A_{S}$ and constant fluid density $\rho = \rho_{0}$, the height $h(t)$ of the interface as a function of time can be derived from Eq.~\ref{eq:surface_flux} as
\begin{equation}  
h(t) = h_0 - \frac{\phi}{\rho_0} t,
\end{equation}
where $h_0$ represents the initial height of the interface. The time for complete evaporation can be calculated as $t_e = \frac{h_0\rho_0}{\phi}$. Here, $\rho_0$ and $h_0$ correspond to the equilibrium values from the lattice Boltzmann (LB) simulation.

In Figure \ref{fig:flat}, we compare LB and analytical results for the normalized height $h/h_0$ of the interface as a function of the normalized time $t^* = t/(h_0\rho_0/\phi)$. We find good agreement between the analytical and LB results for $\Gamma = 0.31$ for $\phi = 0.03, 0.003, 0.0003$, which correspond to interface speeds of approximately $\frac{dh}{dt} \approx 0.03, 0.003, 0.0003$, respectively. The error between the LB simulation and analytical solution grows as the interface height becomes comparable to the interface thickness ($\approx 5$), especially at later stages of evaporation. Thus, we only report results up to $h/h_0 = 0.1$ at $t^* = 0.9$. We report the percentage error in interface position ($h/h_0$) compared to the analytical solution. For $\phi = 0.0003$, the errors are found to be $0.33\%$ and $2.7\%$ at $t^* = 0.52$ and $0.9$, respectively. Similarly, for $\phi = 0.03$, the errors are $0.4\%$ at $t^* = 0.52$ and $3.9\%$ at $t^* = 0.9$. These results suggest that the dependence of $\phi$ on the accuracy of the results is weak, indicating that a single value of $\Gamma$ could be used regardless of the choice of $\phi$.

Furthermore, we demonstrate the effect of $\Gamma$ on the accuracy by including a result with $\Gamma = 0.305$ in Fig.~\ref{fig:flat}. For $\phi = 0.03$ and $\Gamma = 0.305$, the errors are found to be $1.1\%$ at $t^* = 0.52$ and $9.26\%$ at $t^* = 0.9$. $\Gamma = 0.305$ will be used as a common threshold for comparison across all succeeding cases.

\subsubsection{Freely suspended spherical drop} \label{sec:sphere}
A three-dimensional (3D) domain of size $128^{3}$  is used with periodic BC in all directions. The simulation is initialized with a sphere of red fluid of radius $R_{0} = 44$ lattice nodes at the center of the domain and the surroundings with the blue fluid. For a constant mass flux $\phi$ leaving through surface area $A_{S}$ and constant fluid density $\rho = \rho_{0}$, the radius $R(t)$ of the spherical interface as a function of time can be derived from Eq. \ref{eq:surface_flux} as 
\begin{equation}
    R(t) = R_{0} - \frac{\phi}{\rho_{0}} t, \label{eq:sphere_r}
\end{equation}
where $R_{0}$ is the initial radius of the drop. The time to complete evaporation can be found as $t_{e} = \frac{R_{0}\rho_{0}}{\phi} $. $\rho_{0}$ and $R_{0}$ are taken as the equilibrium values from the LB simulation. Since pressure $p$ and density $\rho$ are related in LB via the equation of state (Eq. \ref{eq:EOS}), $\rho$ depends on $R(t)$ because of the Laplace pressure. $R(t)$ can then be derived from Eq.~\ref{eq:surface_flux} taking into account the variable density (shown in Appendix \ref{sec:app1}), leading to     
\begin{equation}
    R(t) + \frac{4 \sigma}{3 c_{s}^2 \rho_{b}} \, ln(R(t)/R_{0}) = - \frac{\phi t}{\gamma \rho_{b}}  + R_{0}. \label{eq:sphere_r2}
\end{equation}

\noindent
The above equation can be solved for $R(t)$ numerically\cite{2020SciPy-NMeth}. We take the density of surrounding (blue) fluid $\rho_{b}=1$, density ratio $\gamma = 1$ and lattice speed of sound $c_{s} = 1/\sqrt{3}$ .

\begin{figure}
\centering
     \begin{subfigure}[b]{0.45\textwidth}
         \centering
         \includegraphics[scale=0.42]{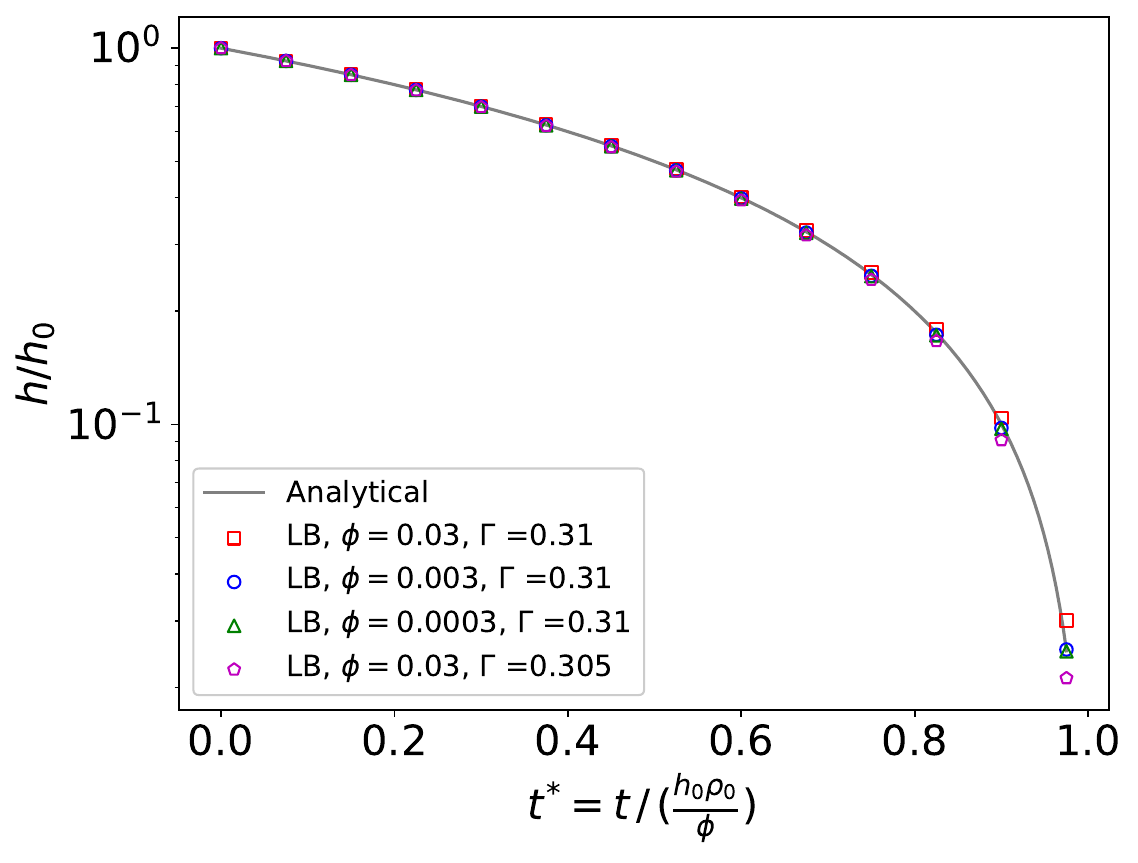}
         \caption{}
         \label{fig:flat}
     \end{subfigure}
     \begin{subfigure}[b]{0.45\textwidth}
         \centering
         \includegraphics[scale=0.42]{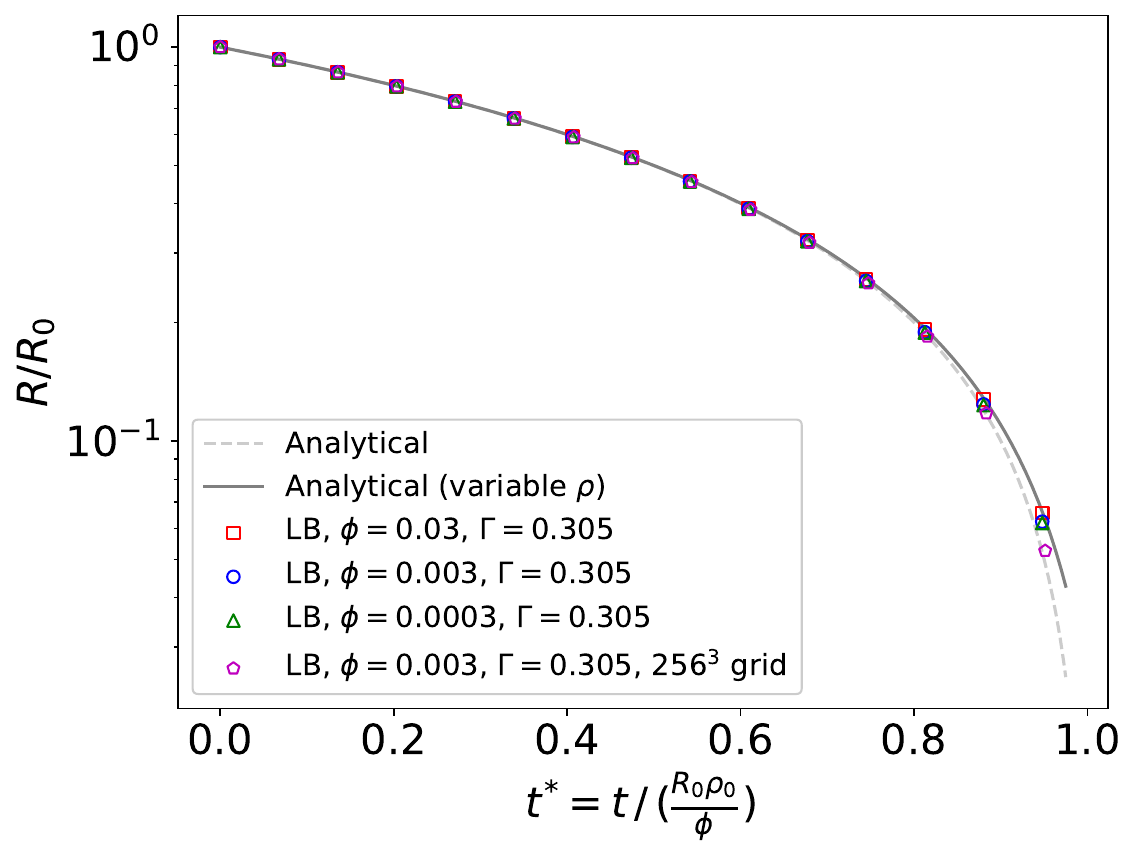}
         \caption{}
         \label{fig:sphere}
     \end{subfigure}
     \begin{subfigure}[b]{0.45\textwidth}
         \centering
         \includegraphics[scale=0.42]{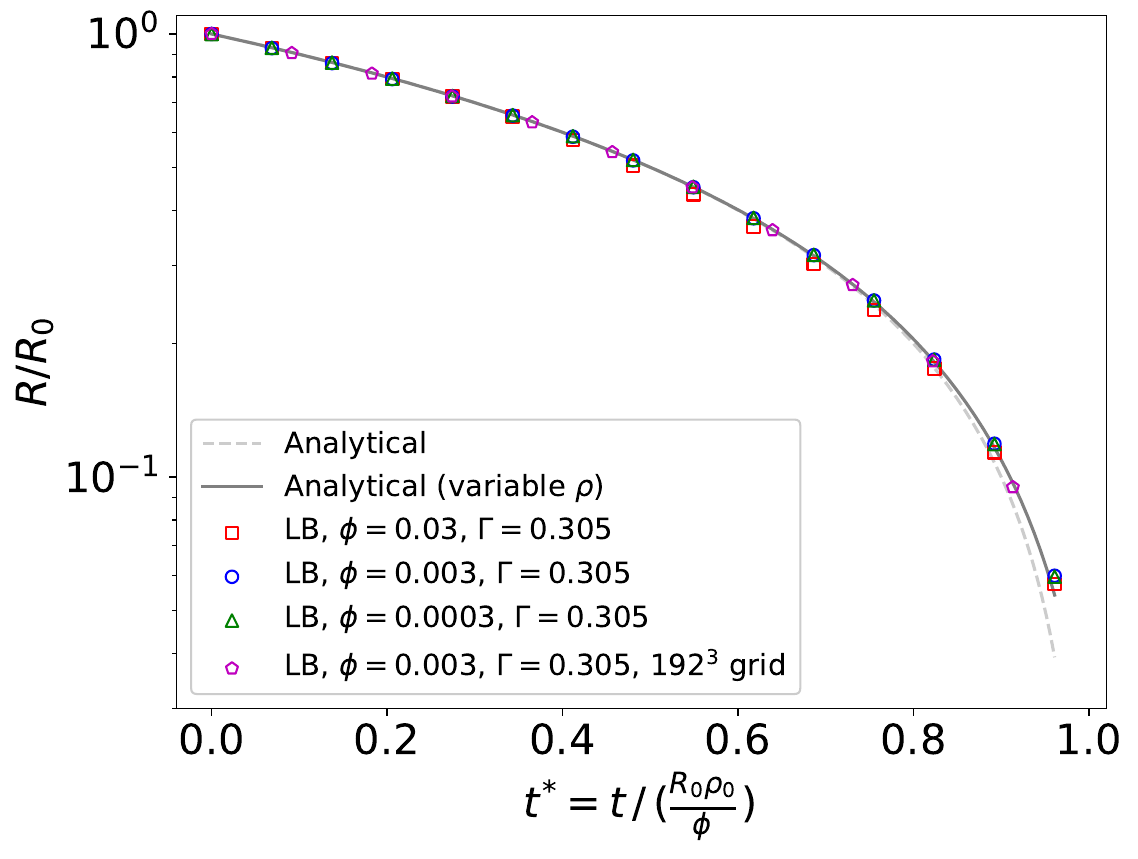}
         \caption{}
         \label{fig:hemisphere}
     \end{subfigure}

\caption{Comparison between LB and analytical results for an evaporating (a) flat interface, (b) spherical drop, and (c) hemispherical drop. Plot of the normalized interface position ($h/h_{0}$ and $R/R_{0}$) with normalized time ($t^{*}$). }
\end{figure}

Figure \ref{fig:sphere} shows the comparison between the LB simulations and analytical results for the normalized drop radius $R/R_{0}$ vs.~the normalized time $t^{*}=t/(R_{0}\rho_{0} /\phi)$. For the analytical results, the effect of variable density is only visible at smaller radii $(R/R_{0}<0.1)$ where the Laplace pressure becomes significant. We compare results till $t^{*}=0.81$ $(R/R_{0}=0.18)$ where the drop radius is still larger than the interface thickness. The error between the analytical and the simulation results is $\approx 3\%$ at $t^{*}=0.81$. Hence, the effect of variable density is insignificant. For $\Gamma = 0.305$, good agreement is found between the analytical (constant density) and LB results for $\phi = 0.03, 0.003, 0.0003$  which correspond to interface speeds of $\frac{dR}{dt} \approx  0.03, 0.003, 0.0003$, respectively. For $\phi = 0.0003$, the error is found to be $0.5 \%$ and $0.4\%$ at $t^{*} = 0.54$ and $0.81$, respectively. For $\phi = 0.03$, the error is found to be $0.2 \%$ and $2.1\%$ at $t^{*} = 0.54$ and $0.81$, respectively. Once again, for the chosen $\Gamma$, the results do not vary significantly with $\phi$. A simulation performed on a $256^{3}$  lattice using $R_{0} = 88$  is also included in Fig.~\ref{fig:sphere} to demonstrate that a single value of $\Gamma$ leads to accurate results regardless of the drop size.

\begin{figure*}
\centering
     \begin{subfigure}[b]{0.45\textwidth}
         \centering
         \includegraphics[scale=0.18]{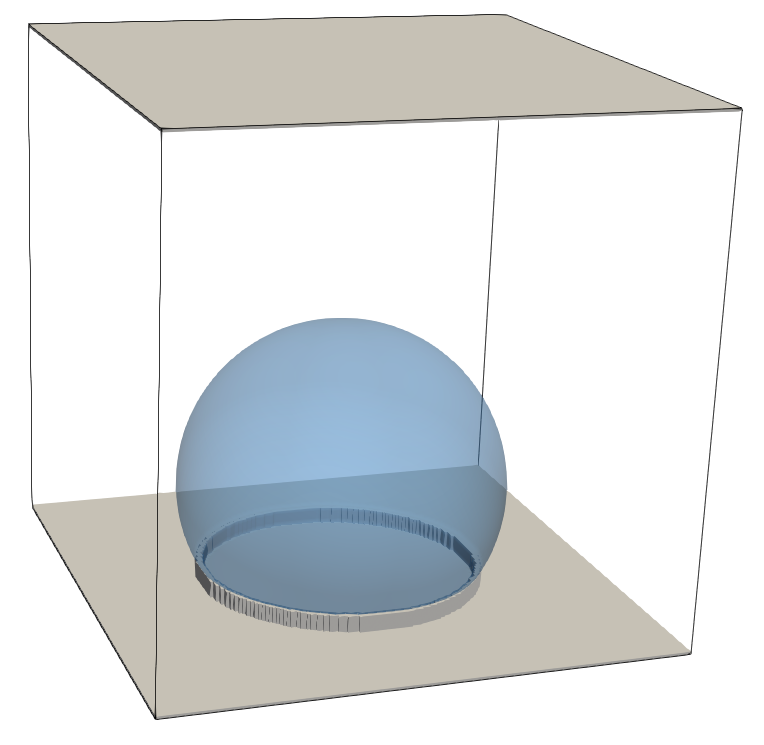}
         \caption{}
         \label{fig:3D_cavity}
     \end{subfigure}
     \begin{subfigure}[b]{0.45\textwidth}
         \centering
         \includegraphics[scale=0.35]{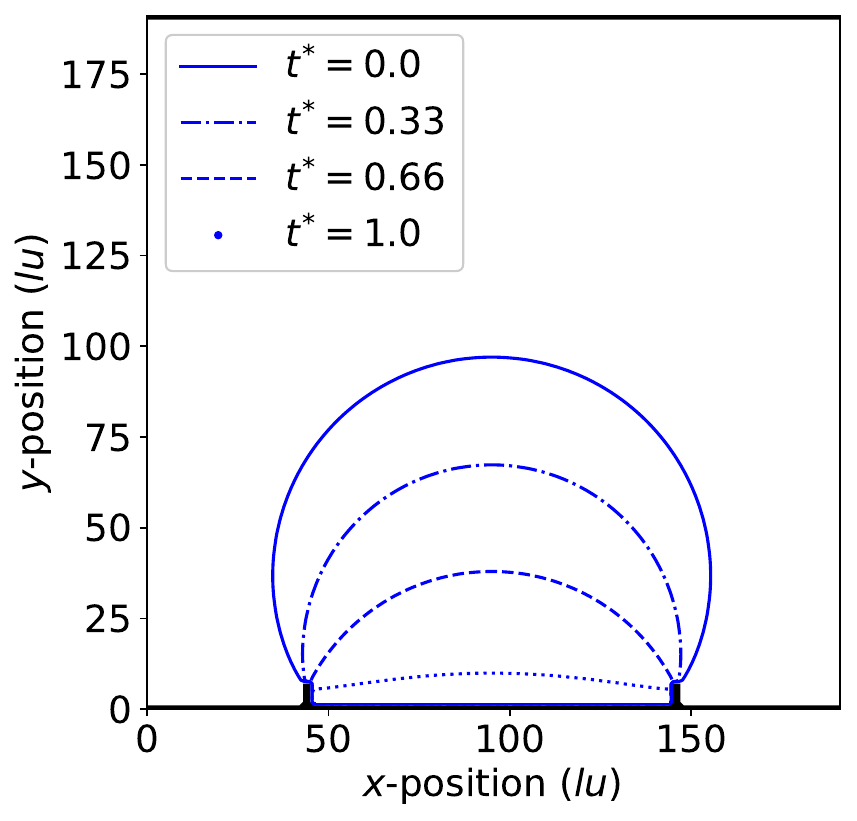}
         \caption{}
         \label{fig:schematic}
     \end{subfigure}
     \begin{subfigure}[b]{0.45\textwidth}
         \centering
         \includegraphics[width=\textwidth]{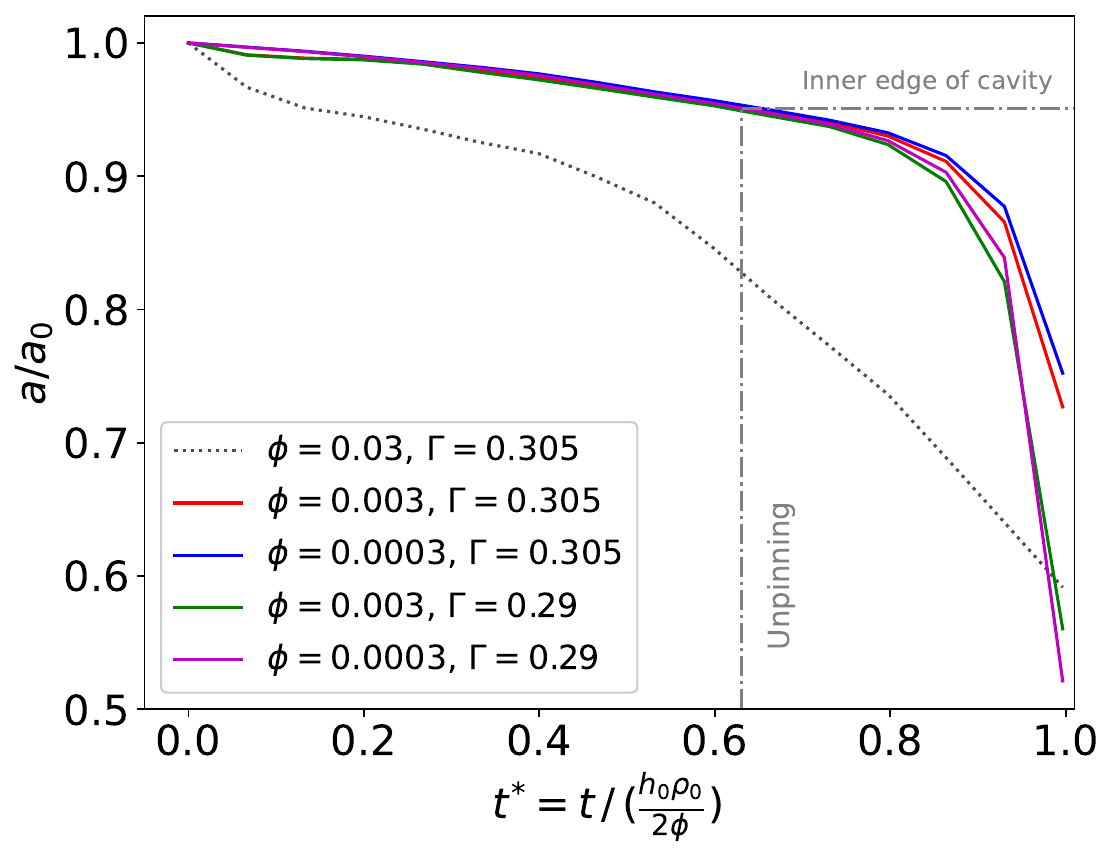}
         \caption{}
         \label{fig:pinned_contact}
     \end{subfigure}
     \begin{subfigure}[b]{0.45\textwidth}
         \centering
         \includegraphics[width=\textwidth]{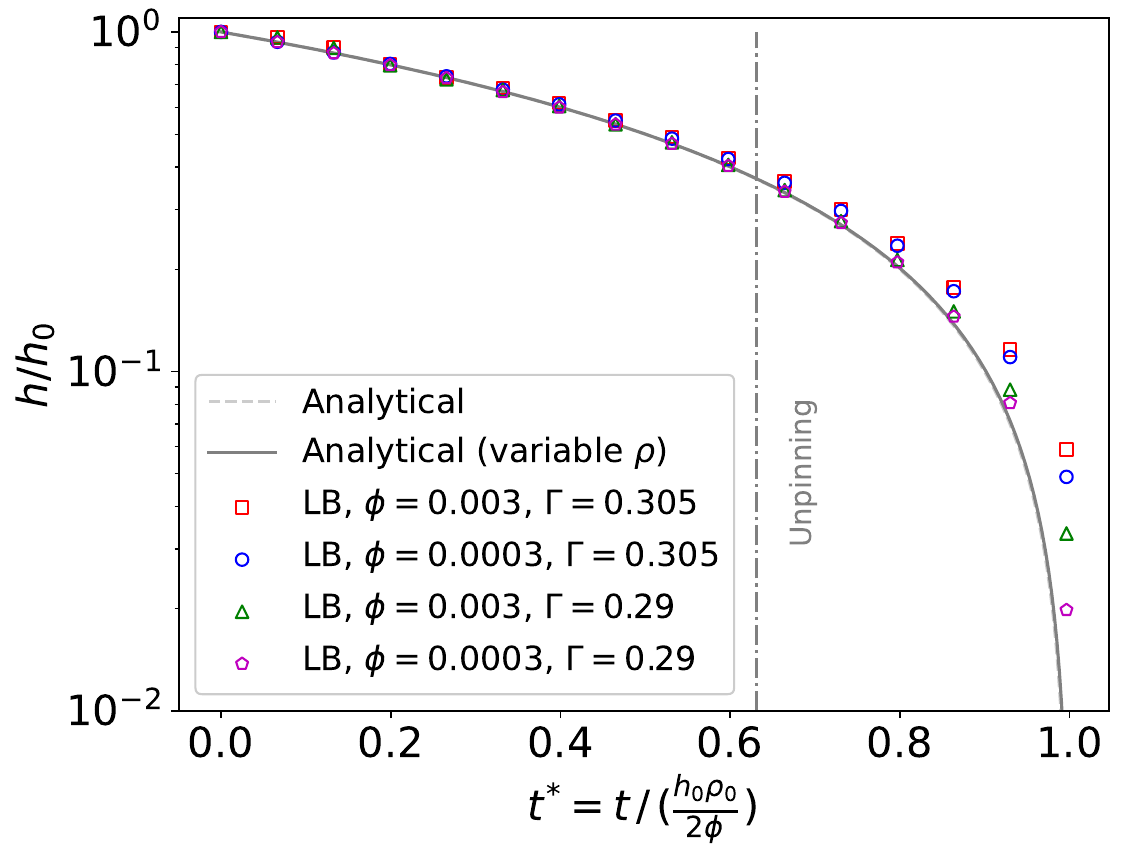}
         \caption{}
         \label{fig:pinned_height}
     \end{subfigure}
\caption{(a) 3D view of pinned spherical drop resting on top of the cylindrical cavity, with walls on the top and bottom of the domain. (b) Drop profile at various stages of evaporation, slice taken at $z$-position midpoint. (c) Plot of the normalized LB contact radius ($a/a_{0}$) with normalized time ($t^{*}$). (d) Comparison between LB and analytical results, plot of the normalized drop height $h/h_{0}$ with normalized time ($t^{*}$).}
\end{figure*}

\subsubsection{Hemispherical drop on substrate} \label{sec:hemisphere}
Similar to the previous section. we use a 3D domain of size $128^{3}$ with periodic boundaries with the exception of no-slip walls at the $y$-direction boundaries. The simulation is initialized with a hemisphere of red fluid of radius $R_{0} = 44$ at the center of the bottom wall. Analytically, the radius $R(t)$ of the hemispherical drop evolves through Eqs. \ref{eq:sphere_r} and \ref{eq:sphere_r2} for constant and variable fluid density, respectively.    

Figure \ref{fig:hemisphere} shows the comparison between LB and analytical results for normalized drop radius $R/R_{0}$ vs. normalized time $t^{*} = t/(R_{0}\rho_{0} /\phi)$. Once again, we find the effect of variable density to be insignificant at $t^{*} = 0.81$, where we compare the results.  For $\Gamma = 0.305$, good agreement is found between the analytical (constant density) and LB results for $\phi = 0.03, 0.003, 0.0003$ which correspond to interface speeds of $\frac{dR}{dt} \approx  0.03, 0.003, 0.0003 $, respectively. For $\phi = 0.0003$, the error is found to be $0.23 \%$ and $4.28\%$ at $t^{*} = 0.54$ and $0.81$, respectively. For $\phi = 0.03$ , the error is found to be $3.6 \%$ and $0.45\%$ at $t^{*} = 0.54$ and $0.81$, respectively. For faster evaporation rates ($\phi = 0.03$), the irregularity in error is observed due to the local contact angle deviating from $90^{0}$ as the contact line moves. As a result, the drop deforms slightly out of its hemispherical shape. Data from a simulation on $192^{3}$ lattice using $R_{0} = 66$ is also included in Fig.~\ref{fig:hemisphere} to demonstrate that the choice of $\Gamma$ can be independent of the drop size.  

\subsubsection{Pinned spherical cap} \label{sec:cap}
A 3D domain of size $192^{3}$  is used with no-slip walls at the $y$-direction boundaries and periodic BC at the remaining boundaries. A cylindrical cavity of inner radius $50$, outer radius $53$, and height $8$ is placed at the center of the bottom wall. A spherical liquid drop is initialized such that its interface is pinned at the outer edge of the cavity at equilibrium, resulting in a contact angle of $\theta \approx 120^{0}$. The setup and equilibrium drop profile are shown in Figs.~\ref{fig:3D_cavity} and
\ref{fig:schematic}.  For a constant mass flux $\phi$ leaving through surface area $A_{S}$ and constant fluid density $\rho = \rho_{0}$, the height $h$ of the spherical cap as a function of time can be derived from Eq.~\ref{eq:surface_flux} as 
\begin{equation}
    h(t) = h_{0} - \frac{2\phi}{\rho_{0}} t, \label{eq:pinned}
\end{equation}
where $h_{0}$ is the initial height of the spherical cap. The time to complete evaporation can be found as $t_{e} = \frac{h_{0}\rho_{0}}{2\phi} $. $\rho_{0}$ and $h_{0}$ (excluding the cavity height) are taken as the equilibrium values from the LB simulation. $h(t)$ can also be derived from Eq.~\ref{eq:surface_flux} taking into account the variable density (shown in Appendix \ref{sec:app2}), leading to   
\begin{eqnarray}
    && h(t) + \frac{4 \sigma}{3 c_{s}^{2} \rho_{b}} \, \left[ ln\left(\frac{a^2 + h(t)^2}{a^2 + h_{0}^2}\right) - \frac{a^4}{(a^2+h(t)^2)^2} \right] = \nonumber \\
    && \frac{- 2 \phi t}{\gamma \rho_{b}} + h_{0} - \frac{4 \sigma}{3 c_{s}^{2} \rho_{b}} \frac{a^4}{(a^2+h_{0}^2)^2}, \label{eq:pinned2}
\end{eqnarray}
where $a$ is the drop contact radius (constant) taken equal to the equilibrium value from the LB simulation ($a = a_{0}$). We take the density of the surrounding (blue) fluid $\rho_{b}=1$ , the density ratio $\gamma = 1$ and the lattice speed of sound $c_{s} = 1/\sqrt{3}$ . Eq.~\ref{eq:pinned2} is solved for $h(t)$ numerically\cite{2020SciPy-NMeth}.

The contact radius is determined by tracking the interface position in a plane just above the cylindrical cavity. As shown in Fig. \ref{fig:pinned_contact}, the contact line moves slowly instead of becoming immobilized. A similar behavior has been observed when pinning the contact line with wetting boundary conditions\cite{li2016pinning}. For the current case, unpinning happens when the contact line reaches the inner edge of the cylindrical cavity ($a/a_{0} = 0.95$) and subsequently starts to move downwards (shown in Fig.~\ref{fig:schematic}). $\phi=0.03$  has not been considered in the results as it leads to a premature unpinning of the contact line, as shown by the dotted line in Fig.~\ref{fig:pinned_contact}.   

Figure \ref{fig:pinned_height} shows the comparison between LB and analytical results for the normalized drop height $h/h_{0}$ vs. normalized time $t^{*} = t/(h_{0}\rho_{0} /2\phi)$. Results are compared till unpinning, which for most cases occurs near $t^{*} \approx 0.63$ (Fig.~\ref{fig:pinned_contact}). For analytical solutions, the difference between constant and variable density results is insignificant owing to the lack of high curvatures. For $\Gamma=0.305$  (the threshold used in previous sections), the error between the LB simulation and the analytical solution at the nearest data point after unpinning (at $t^{*} = 0.664$) is found to be $8\%$ and $7\%$ for $\phi = 0.003$ and $0.0003$, respectively. While at $t^{*} = 0.332$, the error is $2.5\%$ and $1.2\%$ for $\phi = 0.003$ and $0.0003$, respectively  It should be noted that the analytical solution assumes the contact radius to be fixed throughout the evaporation while the same drifts in the simulations (shown in Fig.~\ref{fig:pinned_contact}), hence some error is expected. The same can be offset by adjusting the threshold. For $\Gamma = 0.29$, good agreement is found between the analytical and LB results for $\phi = 0.003$ and $0.0003$  with errors of $1.9\%$ and $0.7\%$ at unpinning, respectively. 

\subsection{Effect of parameters of the color-gradient model} \label{sec:cg_parms}
\subsubsection{Interface thickness parameter}
\begin{figure}
\centering
     \begin{subfigure}[b]{0.45\textwidth}
         \centering
         \includegraphics[scale=0.42]{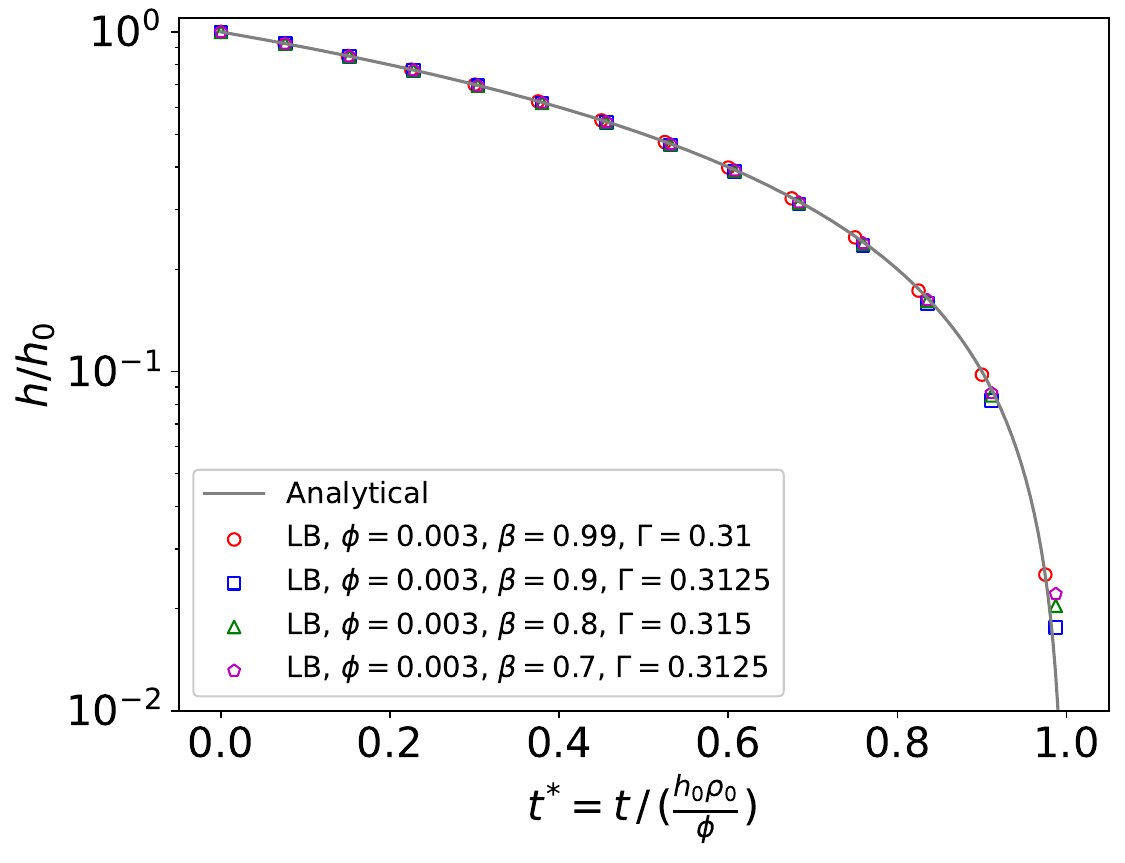}
         \caption{}
         \label{fig:beta_flat}
     \end{subfigure}
     \begin{subfigure}[b]{0.45\textwidth}
         \centering
         \includegraphics[scale=0.42]{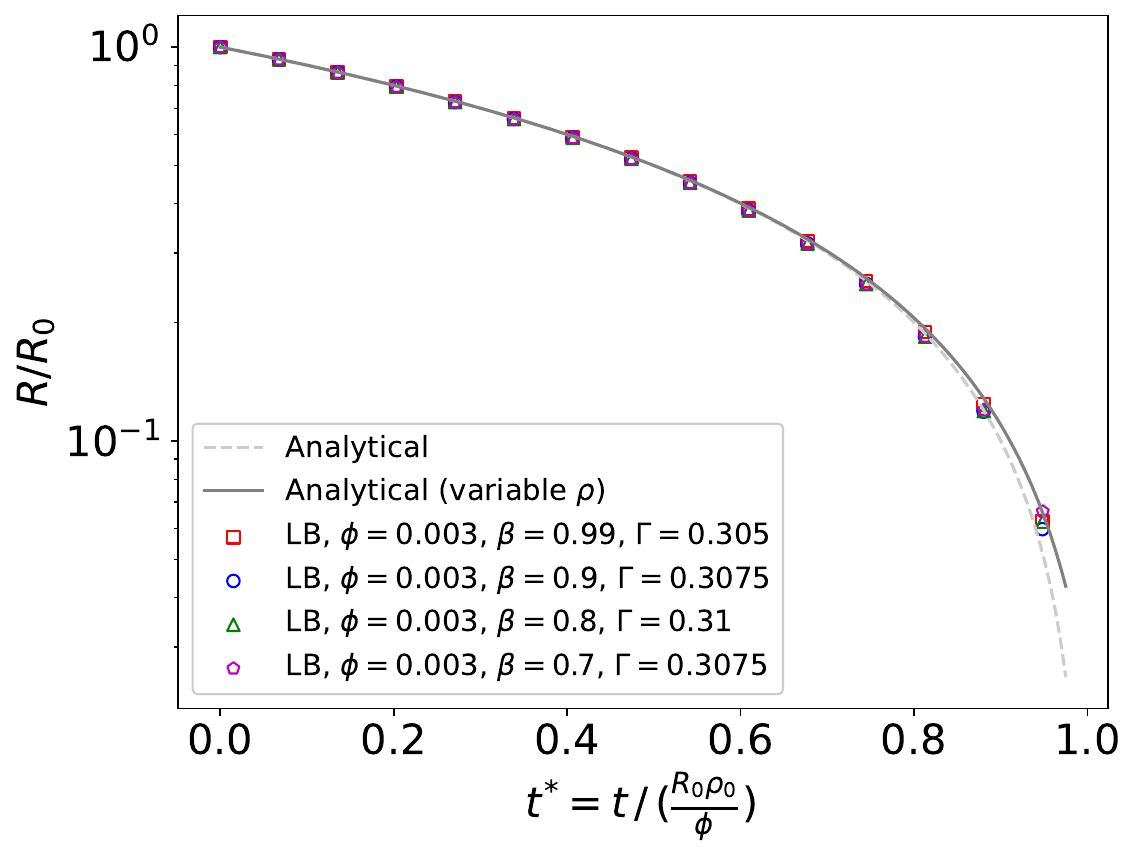}
         \caption{}
         \label{fig:beta_spherical}
     \end{subfigure}
\caption{Comparison between LB simulations and analytical results for an evaporating (a) flat interface and (b) a spherical drop for varying $\beta$. Plot of the normalized interface position ($h/h_{0}$ and $ R/R_{0}$) with normalized time ($t^{*}$).}
\end{figure}

The interface thickness in the CG-LB method is determined by the parameter $\beta$, which ranges from 0 to 1. Here, we present results for $\beta$ values of $0.99$, $0.9$, $0.8$, and $0.7$, corresponding to interface thicknesses of approximately $5$, $6$, $7$ and $8$ lattice nodes, respectively. The benchmarks for flat and spherical interfaces in sections \ref{sec:flat} and \ref{sec:sphere} are repeated with varying $\beta$ in Figs.~\ref{fig:beta_flat} and \ref{fig:beta_spherical}, respectively. It can be observed that the parameter $\Gamma$ varies weakly with the interface thickness. A maximum difference (absolute) of $0.005$ is observed in $\Gamma$ across all the $\beta$ values considered for each of the flat and spherical cases. Additionally, the difference in $\Gamma$ between the flat and spherical cases remains consistent for a given $\beta$ ($|\Gamma_{flat} - \Gamma_{sphere}| = 0.005$). Thus, changing the interface thickness has a minor effect on $\Gamma$, and this effect is uniform across different interface shapes.

\subsubsection{Density contrast}
In the case of a density contrast ($\gamma \neq 1$), an outlet for the non-evaporating fluid component (blue fluid) needs to be added in the domain to maintain the desired density ratio between the two components. This was not necessary for the unit density ratio cases shown previously. We use a pressure boundary condition\cite{Zou_He,Hecht_2010} to implement the outlet for the blue fluid. We replicate the benchmark from Sec.~\ref{sec:sphere} for a spherical interface with a drop radius $R_{0} =44$  and a domain size of $192^{3}$. The simulation is initiated with $\rho_{r} = 2$, $\rho_{b} =1$  for $\gamma = 2$, and $\rho_{r} = 4$, $\rho_{b} =1$  for $\gamma = 4$, followed by equilibration. During the evaporation stage, outlets are specified at each of the $z$-direction boundaries for the blue fluid with $\rho_{b} = 1$ enforced.   

\begin{figure}
\centering
     \begin{subfigure}[b]{0.45\textwidth}
         \centering
         \includegraphics[scale=0.42]{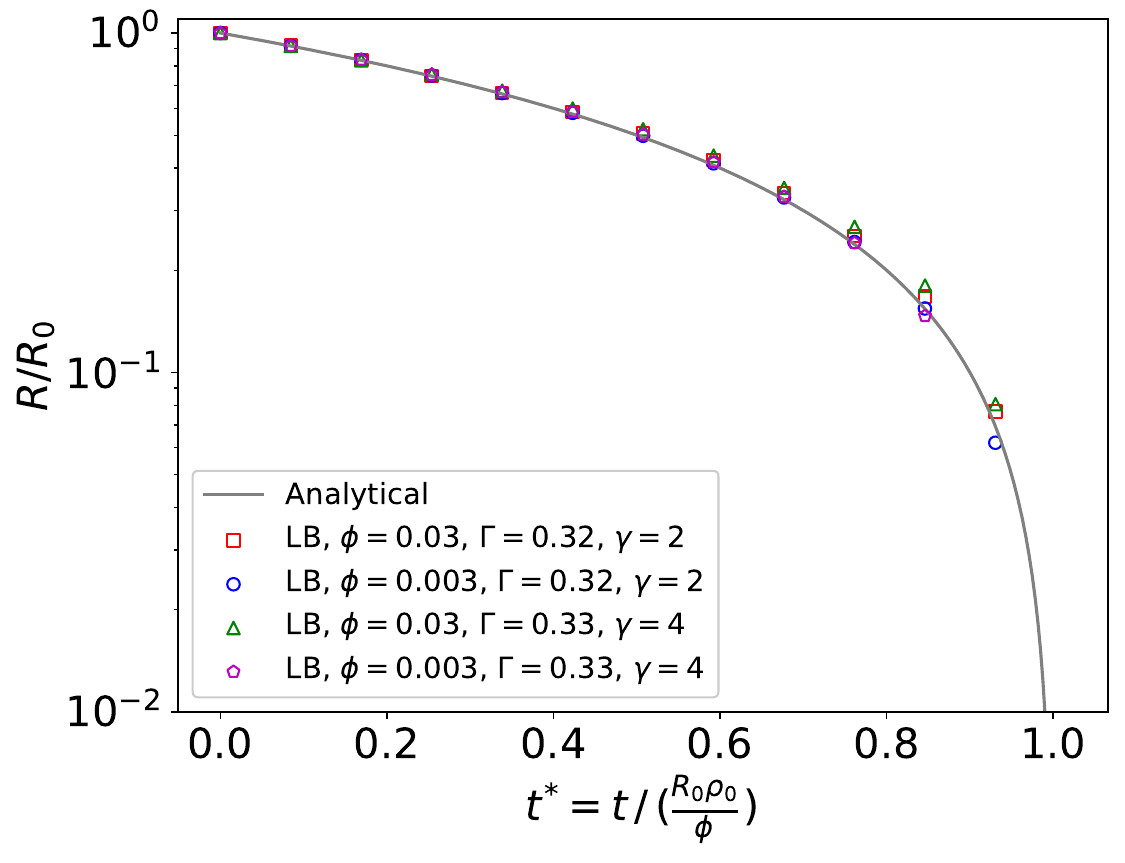}
         \caption{}
         \label{fig:sphere_dens}
     \end{subfigure}
     \begin{subfigure}[b]{0.45\textwidth}
         \centering
         \includegraphics[scale=0.42]{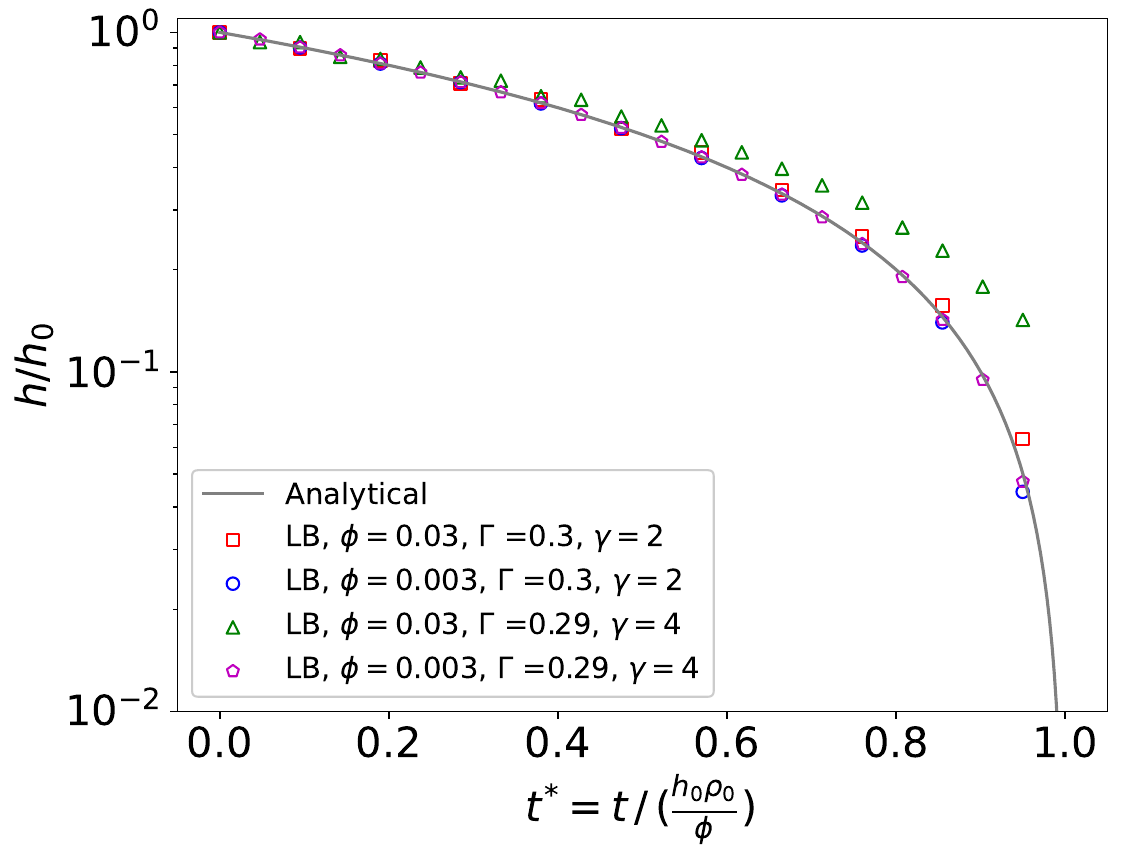}
         \caption{}
         \label{fig:flat_dens}
     \end{subfigure}
\caption{Comparison between LB simulations and analytical results for an evaporating (a) spherical drop and (b) a flat interface for density ratio $\gamma = 2$ and $4$. Plot of the normalized interface position ($h/h_{0}$ and $ R/R_{0}$) with normalized time ($t^{*}$).}
\end{figure}

Figure \ref{fig:sphere_dens} compares LB and analytical (constant density) results for a spherical interface with $\gamma = 2$ and $4$. The following range of errors are reported in the interval $t^{*}=0.5 - 0.85$. For $\gamma = 2$ and $\Gamma=0.32$ , the error is found to be $2.6 - 8.7 \%$ and $0.17 - 1.1\%$ for $\phi = 0.03$ and $0.003$, respectively. For $\gamma = 4$ and $\Gamma=0.33$, the error is found to be $5.7 - 17.1 \%$ and $2 - 4.6\%$ for $\phi = 0.03$ and $0.003$, respectively. In both cases, good agreement is found for smaller $\phi$ while the error increases for larger $\phi$ and higher $\gamma$.

This behavior differs from the observations for $\gamma = 1$, where the accuracy was weakly dependent on $\phi$. This difference stems from the outlet boundary conditions used for the density contrast cases. For smaller $\phi$, the outlet boundary conditions can maintain the blue fluid at the desired density. However, for larger $\phi$, the blue fluid may accumulate over the interface, leading to an increase in density ($\rho_{b}$), which then affects the magnitude of the color-gradient and the accuracy of the results. Depending on the domain geometry and the location/number of outlets, the accuracy can vary significantly when using larger $\phi$.

Figure \ref{fig:flat_dens} recreates the flat interface benchmark from Sec.~\ref{sec:flat} for $\gamma = 2$ and $4$, with an increased domain size of 160$^3$ lattice nodes and an outlet at the upper y-direction boundary. For the case of $\phi = 0.03$  and $\gamma = 4$, we observe the effect described in the previous paragraph more prominently than in the spherical case. The error in the interface position $h/h_{0}$ increases significantly, deviating away from the analytical solution as the evaporation proceeds. Nevertheless, as shown in Fig. \ref{fig:flat_dens}, accurate results can still be achieved with smaller $\phi$ and $\gamma$. 

It is observed that as $\gamma$ increases, the difference in threshold $\Gamma$ between the flat and spherical interfaces also increases: $|\Gamma_{flat} - \Gamma_{sphere}|=$ $0.02$ and $0.04$ for $\gamma=$ $2$ and $4$, respectively. It suggests that $\Gamma$ must be chosen separately for different shapes of the interface to get accurate results for $\gamma > 1$. In contrast, for $\gamma=1$ as shown in Sec. \ref{sec:shape}, a single value of $\Gamma$ gives fairly accurate results across different interface shapes. Hence, the threshold $\Gamma$ needs to be fine-tuned to suit applications involving density contrast and different interface shapes.

\subsection{Evaporation with space and time-varying flux}\label{sec:var_flux}
\begin{figure}
\centering
     \begin{subfigure}[b]{0.45\textwidth}
         \centering
         \includegraphics[scale=0.45]{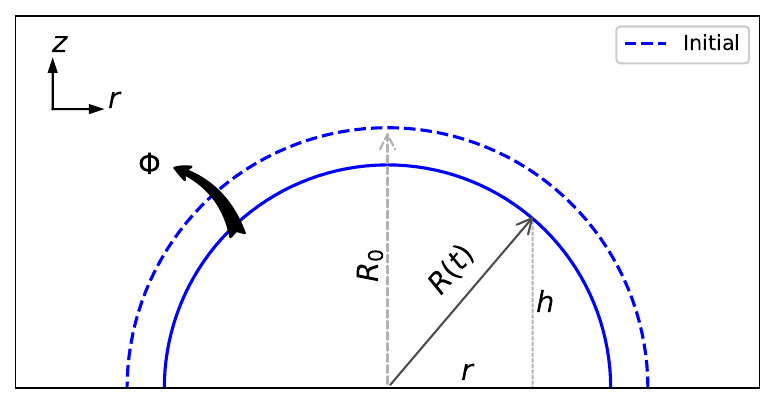}
         \caption{}
         \label{fig:hemi_scheme}
     \end{subfigure}
     \begin{subfigure}[b]{0.45\textwidth}
         \centering
         \includegraphics[scale=0.4]{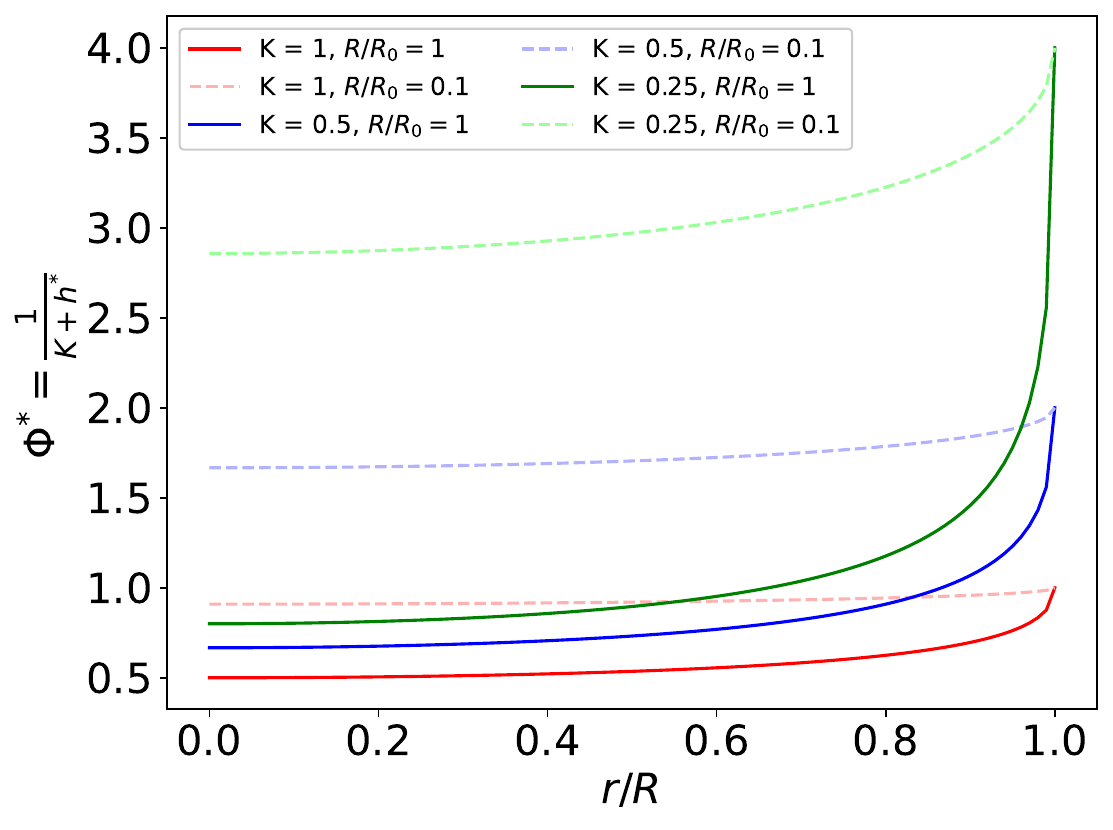}
         \caption{}
         \label{fig:flux_variation}
     \end{subfigure}
     \begin{subfigure}[b]{0.45\textwidth}
         \centering
         \includegraphics[scale=0.4]{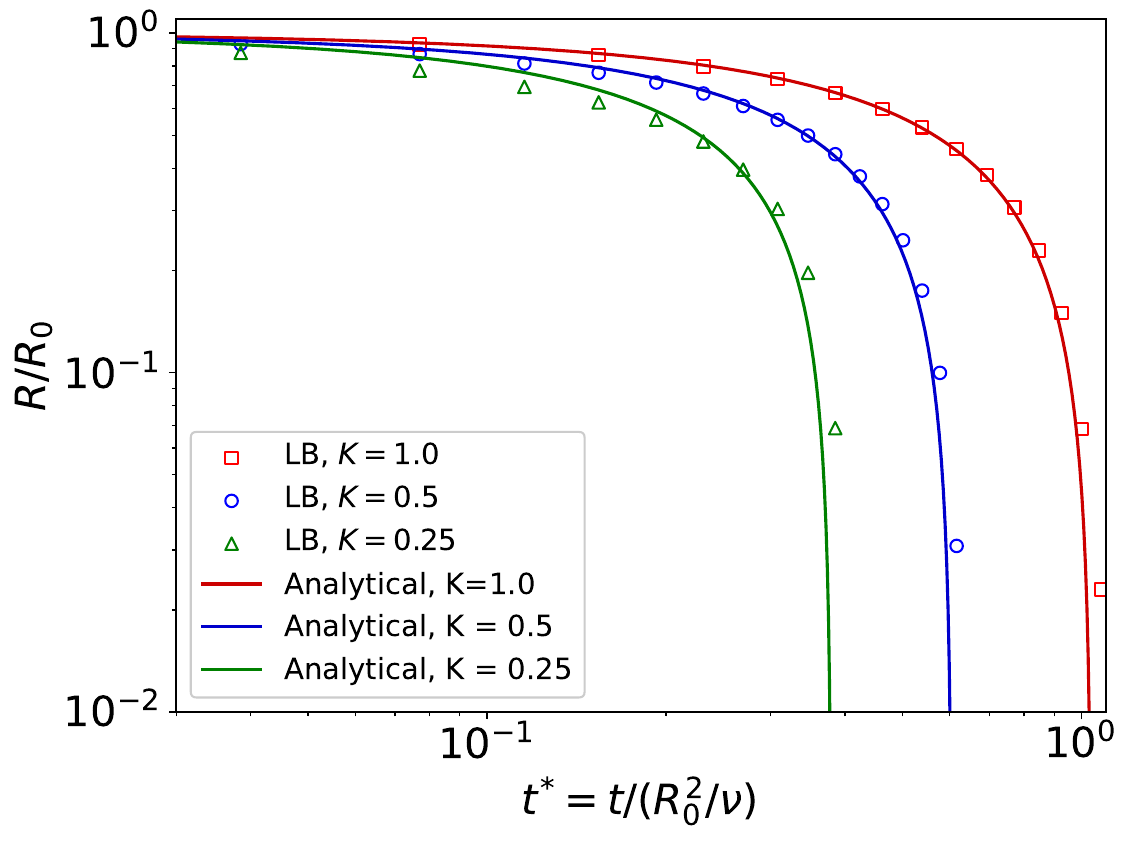}
         \caption{}
         \label{fig:flux_curve}
     \end{subfigure}
\caption{(a) Schematic of hemispherical drop evaporating from a substrate with space and time varying flux $\Phi$ at the interface. (b) Variation of evaporative flux $\Phi^{*}$ with radial positions $r/R$ for different $K$, at the beginning ($R/R_{0}=1$) and nearing the end ($R/R_{0}=0.1$) of evaporation. (c) Comparison between LB and analytical results for variable flux $\Phi$ at different $K$, plot of the normalized drop radius ($R/R_{0}$) with normalized time ($t^{*}$).}
\end{figure}

Unlike the results shown in the previous sections, we now assume a space and time varying evaporative flux. Based on theoretical works of drops evaporating on heated substrates in the reaction-limited regime\cite{MURISIC_KONDIC_2011,pham2017drying,larsson2023comparison}, we adopt a commonly used expression for the flux $\Phi^{*}$ (dimensionless) at the interface, neglecting the Laplace correction and the substrate thickness. It is given by
\begin{equation}
    \Phi^{*} = \frac{1}{K + h^{*}}, \label{eq:heated}
\end{equation}
where $h^{*}$ is the local height of the drop (dimensionless), and $K$ is a constant related to the kinetic effects and volatility of the liquid\cite{larsson2023comparison}. $\Phi^{*}$ varies with drop height as such that the flux is higher near the contact line ($h^{*} \rightarrow 0$) as compared to the top of the drop ($h^{*} \rightarrow 1$).    

We recreate the benchmark from Sec.~\ref{sec:hemisphere} for a hemispherical drop on a $192^{3}$ domain and with $R_{0}=66$ using $\Phi^{*}$. The schematic for the problem depicting relevant parameters is shown in Fig.~\ref{fig:hemi_scheme}. The initial drop radius $R_{0}$ is taken as the length scale, $\phi_{S}$ is taken as the flux scale and $R_{0}^2/\nu$ is taken as the viscous time scale. The dimensional flux can be written as $\Phi= \phi_{S} \, \Phi^{*}$ and $h^{*} = h / R_{0}$. For constant density (assumed for simplicity) and flux $\Phi$, Eq.~\ref{eq:surface_flux} can be solved (shown in Appendix \ref{sec:hemi_drv}) to recover the following differential equation for the drop radius $R(t)$:
\begin{equation}
    \frac{dR(t)}{dt} = \frac{\phi_{S} \, R_{0}}{\rho \, R(t)} \ln\left(\frac{K R_{0}}{R(t) + K R_{0}}\right). \label{eq:drrr}
\end{equation}
Eq.~\ref{eq:drrr} is solved numerically\cite{2020SciPy-NMeth} with initial condition $R(t=0) = R_{0}$. 

For $K = 1, 0.5$ and $0.25$, the variation of $\Phi^{*}$ with the normalized radial coordinate $r/R$ is shown in Fig.~\ref{fig:flux_variation}. Initially ($R/R_{0}=1$), $\Phi^{*}$ increases towards the contact line ($r/R \rightarrow 1$) with the maximum increase for $K = 0.25$ and a minimum increase for $K = 1$. Near the end of the evaporation ($R/R_{0}=0.1$), the variation of flux across the interface diminishes for all $K$ as a smaller height of the drop remains. Hence, $\Phi^{*}$ varies both with space and time. The mass flux scale $\phi_{S}$ is kept fixed at $0.001$. The dimensional flux $\Phi$ can then be converted to a mass sink rate $\varphi$ for LB simulations using Eq.~\ref{eq:relation} at different evaporation sites depending upon the local height $h^{*}$.  

Figure \ref{fig:flux_curve} shows the comparison between LB and analytical results for the normalized drop radius $R/R_{0}$ vs.~the normalized time $t^{*} = t/(R_{0}^{2}/\nu)$ with variable flux $\Phi$ using $K = 1$, $0.5$, and $2.5$. The results presented use the same $\Gamma=0.305$ as in Sec.~\ref{sec:hemisphere}. We report the error at the closest data point to $R/R_{0} \approx 0.2$ (Tab.~\ref{tab:var_flux}). The agreement between LB and analytical results is better for larger $K$, i.e., smaller differences between the maximum and minimum values of flux ($\Phi_{max}/\Phi_{min}$). For $K = 0.5$, an error of $\approx 14 \%$ is observed corresponding to $\Phi_{max}/\Phi_{min} = 3$, which marks the limit of the current method. For $K > 1$, the flux does not vary strongly with space and time ($\Phi_{max}/\Phi_{min} \approx 1$), and the error is expected to be much lower (not shown here). Therefore, the current method can be used with weakly space-time varying flux with reasonable error. It is to be noted that with a larger evaporation rate close to the contact line, the drop can slightly deform out of the hemispherical shape, leading to some error in the early stages of evaporation (visible for $K=0.25$ in Fig.~\ref{fig:flux_curve}). 

\begin{table}
\caption{\label{tab:var_flux} Reported error in interface position for an evaporating hemispherical drop with variable flux $\Phi$ for different $K$, with $\Gamma = 0.305$.}
\begin{ruledtabular}
\begin{tabular}{lccr}
$K$  &   $\Phi_{max}/\Phi_{min}$      & Error &   At $R/R_{0}$  \\
\hline
1 & 2 & 6.71 \% & 0.23   \\ 
0.5 & 3 & 14.39 \% & 0.21   \\ 
0.25 & 5 & 43.83 \% & 0.2   \\ 
\end{tabular}
\end{ruledtabular}
\end{table}

\subsection{Method overview} \label{sec:strategy}
We summarize the findings from the previous sections and comment on the optimum strategy for choosing $\Gamma$. Table \ref{tab:fbar} shows the error in interface position for different shapes considered in Section \ref{sec:shape} when a common $\Gamma=0.305$ is used and for the largest $\phi$ possible. A maximum error of approximately $9\%$ is observed for the flat interface, while it is under $8 \%$ for the curved interfaces. The upper range of the error is generally observed at the end of evaporation and is not indicative of the majority duration, for which the error is usually well under $5\%$. The interface shapes considered to span from no curvature (flat) to increasing curvature (spherical and hemispherical) to decreasing curvature (pinned spherical cap). Different drop evaporation modes, namely, constant contact angle and constant contact radius, are also taken into account in the hemispherical and pinned spherical cap cases. Therefore, a single value of $\Gamma$ could be used for a general shape of the interface with reasonable error for unit density ratio and associated set of CG model parameters. The optimum value of $\Gamma$ could be obtained by recreating some of the benchmarks detailed in Sec.~\ref{sec:shape} for flat and curved interfaces. If the shape of the interface and its evolution are previously known, then the $\Gamma$ could be adjusted for better accuracy via comparison with analytical or numerical results.

From the results in Sec.~\ref{sec:cg_parms}, $\Gamma$ is found to be mainly dependent on the  interface thickness and density ratio. However, the effect of changing the interface thickness on $\Gamma$ was observed to be uniform across different shapes, thus maintaining generality. For $\gamma = 1$, the method is shown to work reliably over a wide range of evaporative flux and interface shapes. For $\gamma \neq 1$, the method requires an outlet for the non-evaporating fluid component to maintain the desired density ratio. Accurate results are observed when both $\phi$ and $\gamma$ are kept small. Additionally, it is recommended to tune $\Gamma$ on a case-by-case basis for accurate results.     

While the current model for RL evaporation is derived for a constant flux $\phi$ (Sec.~\ref{sec:rl_model}), the results from Sec.~\ref{sec:var_flux} demonstrate that it can be used with space and time-varying flux as long as the variation in the flux is not large ($\Phi_{max}/\Phi_{min} \leq 3$). For constant flux and unit density ratio, the method has shown good accuracy over two orders of magnitude of evaporative flux ($\phi = 0.01 - 0.0001$), covering both the fast and slow evaporation rates. The maximum rate is limited by the fluid density of the evaporating fluid component ($\rho_{r}$) and the fluid interface speed being less than $0.1$  (to avoid compressibility errors).       

The dependence of $\Gamma$ on $\bar{S}$ was shown in Figs.~\ref{fig:S_3}, \ref{fig:S_2} and \ref{fig:S_fail}. $\bar{S}>1$ is recommended to avoid holes in the selected evaporation sites around the interface. Additionally, $\bar{S}=3$ was used in the current study due to the minimal dependence of $\phi$ on the accuracy of the results for a given $\Gamma$. The same dependence was found to increase for $\bar{S} < 3$. For instance between $\phi = 0.03$ and $0.003$, the error is found to vary by approximately $1.5 \%$, $9.5 \%$ and $13 \%$ for $\bar{S} = 3$, $2$ and $1$, respectively, for the spherical interface case (Sec.~\ref{sec:sphere}).    

\begin{table}
\caption{\label{tab:fbar} Reported error in interface position for cases considered in Section \ref{sec:shape} for $\Gamma=0.305$.}
\begin{ruledtabular}
\begin{tabular}{lcr}
Interface shape  &          & Error in interface position\footnote{reported at the largest $\phi$ wherever possible} \\
\hline
Flat              &   Section \ref{sec:flat}           &        1.1 - 9.26 \%        \\
Spherical         & Section \ref{sec:sphere}           &        0.2 - 2.1 \%            \\
Hemispherical     & Section \ref{sec:hemisphere}       &        0.45 - 3.6 \%                 \\
Pinned spherical cap & Section \ref{sec:cap}           &        2.5 - 8 \%                \\ 

\end{tabular}
\end{ruledtabular}
\end{table}

\section{Conclusion} \label{sec:end}
We have developed a method for achieving reaction-limited evaporation for the CG LB multicomponent model and have successfully validated it against analytical solutions for various interface shapes. This method involves tuning a single free parameter (threshold) to select evaporation sites at the interface and then exchanging fluid mass between the components. We have demonstrated that a single threshold value can be used for a general shape of the interface with reasonable error over a wide range of flux magnitudes for a given set of CG model parameters, especially for the unit density ratio. However, for density contrast, we recommend using smaller evaporation fluxes and density ratios for accurate results.

Our method utilizes the inherently calculated color-gradient magnitude and the resting populations for mass exchange, making it computationally inexpensive, locally applicable, and easy to implement. Importantly, it does not require any changes to the core algorithm of CG, making it compatible with other CG variants as well. Similar to the NEOS models used in theoretical studies, evaporation occurs solely at the interface and is decoupled from the vapor phase dynamics.

The current method with constant evaporation flux can be used to study isothermal RL evaporation, for example, when there is a neutral gas draft over the interface. The flux can be modeled by the Hertz-Knudsen relation as done by Hernandez et al.\cite{hernandez2017simple} for the evaporation of drops from millimeter-sized pillars, showing good agreement with experimental results. The current method can also be used to study non-isothermal RL evaporation if the flux varies weakly with space and time. For drop evaporation on heated substrates the experimental value for the $K$ parameter in the space-time varying flux (Eq.~\ref{eq:heated} used in Sec.~\ref{sec:var_flux}) is $K = 10$ for water and $K = 1$ for isopropanol as suggested by Murisic and Kondic \cite{MURISIC_KONDIC_2011}. As a result, the flux varies weakly with the drop height for these fluids, and the current method can be used with reasonable error for such applications.  The method also extends the CG model to study evaporation in porous media with possible applications in soil water evaporation\cite{li2019evaluation} and polymer electrolyte fuel cells\cite{safi2017experimental}.

\begin{acknowledgments}
We thank Marcello Sega and Johannes Hielscher for fruitful discussions and technical support. We gratefully acknowledge the financial support provided by the Deutsche Forschungsgemeinschaft (DFG) through the research unit FOR2688 ‘Instabilities, Bifurcations and Migration in Pulsatile Flows’ (Project-ID 349558021) and the collaborative research centers SFB1411 (Project-ID 416229255) and SFB1452 (Project-ID 431791331).
\end{acknowledgments}

\section*{Author Declarations}
\subsection*{Conflict of Interest}
The authors have no conflicts to disclose.

\subsection*{Author Contributions}
\textbf{Gaurav Nath}: Conceptualization (equal); Data curation (equal);
Formal analysis (equal); Investigation (equal); Software (equal); Validation (equal); Writing – original draft (equal). \textbf{Othmane Aouane}:
Conceptualization (equal); Validation (equal); Writing – review \&
editing (equal). \textbf{Jens Harting}: Conceptualization (equal); Funding
acquisition (lead); Writing – review \& editing (equal).

\section*{Data Availability Statement}
The data that support the findings of this study are openly available in Zenodo at http://doi.org/10.5281/zenodo.13382917.

\appendix

\section{$R(t)$ for a spherical drop with variable density} \label{sec:app1}
For a pair of fluid components ($\rho_{r}$, $\rho_{b}$) with density ratio $\gamma = \rho_{r}/\rho_{b} \; (\rho_{r} > \rho_{b})$ and surface tension $\sigma$, the pressure $p$ in each component is given by (using Eq. \ref{eq:EOS})
\begin{eqnarray}
    p_{b}  = \rho_{b} \zeta [1 - W_{0}] &= \rho_{b} \, c_{s}^{2} \label{eq:pr_a},\\ 
    p_{r}  = \rho_{r} \zeta \frac{[1 - W_{0}]}{\gamma} &= \frac{\rho_{r} \, c_{s}^{2}}{\gamma}.\label{eq:pr_b} 
\end{eqnarray}
Assuming red fluid inside the spherical drop and blue fluid surrounding it, according to Young-Laplace equation we have
\begin{equation}
    p_{r} - p_{b} = \frac{2 \sigma}{R}.
\end{equation}
Substituting the pressures from Eqs.~\ref{eq:pr_a} and \ref{eq:pr_b} in the above expression, we get
\begin{equation}
     \rho_{r} = \gamma \, \left[\rho_{b} + \frac{2 \sigma}{R c_{s}^{2}} \right]. \label{eq:rho_in}
\end{equation}
Here, we assume $\rho_{b}$ to be a constant. For a constant mass flux $\phi$ leaving through surface area $A_{S}$ and variable density $\rho(t)$, Eq. \ref{eq:surface_flux} can be written as 
\begin{eqnarray}
    \frac{d(\rho V)}{dt} &=& - \phi A_{S}, \\
    \frac{dV}{dt} &=& -\frac{\phi}{\rho} A_{S} - \frac{V}{\rho} \frac{d\rho}{dt}. \label{eq:varflux}
\end{eqnarray}
Substituting the volume $V = (4/3)\pi R^{3}$, surface area $A_{S} = 4 \pi R^{2}$ and Eq.~\ref{eq:rho_in} for $\rho$ in the above equation leads to
\begin{equation}
    \frac{dR}{dt} = -\frac{\phi / \gamma }{\frac{4 \sigma}{3 R c_{s}^{2}}+\rho_{b}}. \label{eq:diff_eq}
\end{equation}
Solving the above differential equation using the initial condition $R = R_{0}$ at $t = 0$, we get
\begin{equation}
     R(t) + \frac{4 \sigma}{3 c_{s}^2 \rho_{b}} \, ln(R(t)/R_{0}) = - \frac{\phi t}{\gamma \rho_{b}}  + R_{0}.
\end{equation}
This equation is solved for $R(t)$ numerically\cite{2020SciPy-NMeth}.

\section{$h(t)$ for a pinned spherical drop with variable density} \label{sec:app2}
For a spherical cap with contact radius $a$ and height $h$, the volume $V$ and surface area $A_{S}$ of the spherical cap are given by $V = \frac{1}{6} \pi h (3 {a}^2 + {h}^2) $ and $A = \pi ({a}^2 + {h}^2)$, respectively. The density of the drop $\rho$ is given by Eq.~\ref{eq:rho_in}, where $R = \frac{a^{2} + h^{2}}{2 h}$ is the radius of curvature of the spherical cap. Substituting the volume $V$, surface area $A_{S}$ and $\rho$ in Eq.~\ref{eq:varflux} and simplifying for $h(t)$, we get
\begin{equation}
    \frac{d h}{d t} = \frac{-2 \phi / \gamma}{\frac{4 \sigma h}{(a^{2}+h^{2}) c_{s}^2} + \rho_{b} + \frac{4 \sigma}{3 c_{s}^2} \, h \, (3 a^2 + h^2) \frac{a^2 - h^2}{(a^2 + h^2)^{3}}}, 
\end{equation}
where the contact radius $a$ is taken as constant for the pinned spherical cap. Solving the above differential equation using the initial condition $h = h_{0}$ at $t = 0$, we get
\begin{eqnarray}
    && h(t) + \frac{4 \sigma}{3 c_{s}^{2} \rho_{b}} \, \left[ ln\left(\frac{a^2 + h(t)^2}{a^2 + h_{0}^2}\right) - \frac{a^4}{(a^2+h(t)^2)^2} \right] = \nonumber \\ && \frac{- 2 \phi t}{\gamma \rho_{b}} + h_{0} - \frac{4 \sigma}{3 c_{s}^{2} \rho_{b}} \frac{a^4}{(a^2+h_{0}^2)^2}. 
\end{eqnarray}
The above equation is solved for $h(t)$ numerically\cite{2020SciPy-NMeth}.

\section{$R(t)$ for a hemispherical drop with space and time varying flux}\label{sec:hemi_drv}
We choose cylindrical coordinates ($r,\theta, z$) for the current case with the problem schematic and relevant parameters shown in Fig.~\ref{fig:hemi_scheme}. The initial drop radius $R(t=0) = R_{0}$ is chosen as the length scale. For a hemispherical drop, the drop height at any given $r$ is given by $h = \sqrt{R(t)^2 - r^2}$ and the normalized height is given by $h^{*} = h/R_{0}$. 

Let $\Phi^{*}$ be the space- and time-varying flux (non-dimensional) at the interface dependent on the local drop height $h^{*}$, as given by
\begin{equation}
    \Phi^{*} = (K + h^{*})^{-1} = \left(K + \frac{\sqrt{R(t)^2 - r^2}}{R_{0}} \right)^{-1}, \label{eq:Phi_star}
\end{equation}
where $K$ is a constant. $\Phi^{*}$ can be written in a dimensional form assuming a flux scale $\phi$, such that $\Phi = \phi_{S} \, \Phi^{*}$.

Eq.~\ref{eq:surface_flux} for constant $\rho$ and variable flux $\Phi = \phi_{S} \, \Phi^{*}$ can be written as
\begin{equation}
    \frac{dV}{dt} = \frac{-\phi_{S}}{\rho} \iint_{A_{S}} \Phi^{*} \, dA_{S}. \label{eq:var_flux_}
\end{equation}
In cylindrical coordinates, the elemental surface area on the hemisphere is given by
\begin{equation}
    dA_{S} = \frac{R } {\sqrt{R^2 - r^2}}  r \, dr \, d\theta.
\end{equation}
Substituting $dA_{S}$ from the above equation, $V =\frac{2}{3}\pi R^{3}$ and $\Phi^{*}$ from Eq.~\ref{eq:Phi_star} in Eq.~\ref{eq:var_flux_},  we get  
\begin{eqnarray}
    \frac{d}{dt}\left( \frac{2}{3}\pi R^{3} \right) = && 
    - \frac{\phi_{S}}{\rho} \int_{0}^{2\pi} \int_{0}^{R}  \left(K + \frac{\sqrt{R(t)^2 - r^2}}{R_{0}} \right)^{-1} \nonumber \\ && \frac{R}{\sqrt{R^2-r^2} }   r \, dr \, d\theta
\end{eqnarray}
After simplifying, we then get
\begin{equation}
    \frac{dR(t)}{dt} = \frac{\phi_{S} \, R_{0}}{\rho \, R(t)} \ln\left(\frac{K R_{0}}{R(t) + K R_{0}}\right). 
\end{equation}
The above differential equation is solved for $R(t)$ numerically via an explicit Runge-Kutta method\cite{2020SciPy-NMeth} of order 5(4) with initial condition $R(t=0) = R_{0}$.  

\end{document}